\documentclass[aps,prd,reprint,onecolumn, groupedaddress,longbibliography, 12pt]{revtex4-1}
\usepackage{amsmath,amssymb}
\usepackage{graphicx}
\usepackage{color}
\usepackage{caption}
\usepackage{hyperref}
\begin{document}
%------------------------------------------------------------------------------------------------------------------------------------------
% very standard definitions
%------------------------------------------------------------------------------------------------------------------------------------------
\def\d{{\mathrm{d}}}
\newcommand{\scri}{\mathscr{I}}
\newcommand{\sun}{\ensuremath{\odot}}
\def\J{{\mathscr{J}}}
\def\L{{\mathscr{L}}}
\def\sech{{\mathrm{sech}}}
\def\T{{\mathcal{T}}}
\def\tr{{\mathrm{tr}}}
\def\diag{{\mathrm{diag}}}
\def\ln{{\mathrm{ln}}}
%------------------------------------------------------------------------------------------------------------------------------------------
\def\Horava{Ho\v{r}ava}
\def\Aether{\AE{}ther}
\def\AEther{\AE{}ther}
\def\aether{\ae{}ther}
\def\UH{{\text{\sc uh}}} % small caps
\def\KH{{\text{\sc kh}}} % small cpas
\def\ks{{ k_s }}
%------------------------------------------------------------------------------------------------------------------------------------------
% very standard definitions
%------------------------------------------------------------------------------------------------------------------------------------------
\def\n{\mathbf{n}}
\def\x{\mathbf{x}}
\def\s{\mathbf{s}}
%------------------------------------------------------------------------------------------------------------------------------------------
\title{Ray tracing Einstein--{\AE}ther black holes: \\[5pt] Universal versus Killing horizons\\[10pt]}
%\title{Ray Trajectories for Einstein--{\AE}ther black holes: \\[5pt] Universal versus Killing horizons\\[10pt]}
\author{Bethan Cropp}
\email[]{bcropp@sissa.it}
\author{Stefano Liberati}
\email[]{stefano.liberati@sissa.it}
\author{Arif Mohd}
\email[]{arif.mohd@sissa.it}
\affiliation{SISSA, Via Bonomea 265, 34136 Trieste, Italy
                                       and \\
     INFN sezione di Trieste, Via Valerio 2, 34127 Trieste, Italy.}
\author{Matt Visser}
\email[]{matt.visser@msor.vuw.ac.nz}
\affiliation{ \mbox{School of Mathematics, Statistics, and Operations Research,}\\
Victoria University of Wellington; \\
PO Box 600, Wellington 6140, New Zealand.\\}
%\thanks{}
\date{21 November 2013; \LaTeX-ed \today}
%------------------------------------------------------------------------------------------------------------------------------------------

\begin{abstract}

\smallskip
%\noindent
Violating Lorentz-invariance, and so implicitly permitting some form of super\-luminal communication, necessarily alters the notion of a black hole. Nevertheless, in both Einstein-{\AE}ther gravity, and \Horava--Lifshitz gravity, there is still a causally disconnected region in black-hole solutions; now being bounded by a ``Universal horizon'', which traps excitations of arbitrarily high velocities. 
To better understand the nature of these black holes, and their Universal horizons, we study ray trajectories in these spacetimes. We find evidence that Hawking radiation is associated with the Universal horizon, while the ``lingering'' of ray trajectories near the Killing horizon hints at reprocessing there. In doing this we solve an apparent discrepancy between the surface gravity of the Universal horizon and the associated temperature derived by tunneling method. 
These results advance the understanding of these exotic horizons, and provide hints for a full understanding of black-hole thermodynamics in Lorentz-violating theories. 

\bigskip

\end{abstract}
%------------------------------------------------------------------------------------------------------------------------------------------
\pacs{}
%------------------------------------------------------------------------------------------------------------------------------------------
\maketitle
%------------------------------------------------------------------------------------------------------------------------------------------
\clearpage
%------------------------------------------------------------------------------------------------------------------------------------------

\bigskip
\hrule
\tableofcontents
\bigskip
\hrule

%------------------------------------------------------------------------------------------------------------------------------------------

\clearpage
%------------------------------------------------------------------------------------------------------------------------------------------
\section{Introduction}
%------------------------------------------------------------------------------------------------------------------------------------------
\parskip 3 pt
%------------------------------------------------------------------------------------------------------------------------------------------

In spite of being a cornerstone of modern physics, and a very well tested symmetry of nature up to very high energies~\cite{Liberati:2013xla}, Lorentz invariance has been subject to an intense observational and experimental scrutiny in recent years. 
We do not yet have a complete understanding of the behaviour of spacetime at the Planck scale, but many preliminary investigations seem to hint that if spacetime emerges from some discrete substratum of fundamental objects then also its defining symmetries, such as Local Lorentz Invariance (LLI), could be just accidental symmetries at low energies but  broken at high energies. 
Also, recent developments have shown that Lorentz-violating theories of gravity might allow a more manageable UV behaviour, and be at least power-counting renormalizable~\cite{Horava:2009uw,Horava:2008ih,Visser:2009fg}.  

In this sense a strong inspirational role has been played by the Analogue Spacetime framework (also called Analogue Gravity, see for example~\cite{Barcelo:2005fc}), which provided a very large number of toy models for emergent spacetime and emergent local Lorentz invariance. 
For example, under suitable conditions linearized perturbations on an acoustic flow propagate as a scalar field on a curved spacetime. At long wavelengths the perturbations have a ``relativistic'' dispersion relation with the speed of sound as an effective limit speed. 
This ``analogue Lorentz invariance" is violated at short wavelengths as the latter are able to probe the underlying discrete substratum provided by the molecular structure. 
Shaping the flows in these systems allows one to reproduce several geometries and, in particular analogue black-hole solutions, and their predicted thermodynamic properties. In turn, these analogue black holes provide a setup for testing the robustness of these properties against the aforementioned departure from Lorentz invariance in the UV.

This stream of investigation has led to some interesting discoveries, for example, it has been shown that Hawking radiation is robust against the violation of LLI in the UV. 
One also stumbles upon a serious problem though: Violation of LLI seems to lead to violation of the Generalized Second Law (GSL). It is found that if different fields on a black hole spacetime are endowed with different limit speeds, then in general they will experience different horizons and ergoregions. 
This mismatch can then be used to generate violation of the second law, for example, by constructing a {\em perpetuum mobile}, or by devising a setup whose only outcome is to lower the total entropy of the universe~\cite{Dubovsky:2006vk,Eling:2007qd, Jacobson:2008yc}.

A famous quote of Sir Arthur Stanley Eddington (The Nature of the Physical World (1915), chapter 4) says: {\em ``
%The law that entropy always increases holds, I think, the supreme position among the laws of Nature. 
If someone points out to you that your pet theory of the universe is in disagreement with Maxwell's equations -- then so much the worse for Maxwell's equations. If it is found to be contradicted by observation -- well, these experimentalists do bungle things sometimes. 
But if your theory is found to be against the second law of thermodynamics I can give you no hope; there is nothing for it but to collapse in deepest humiliation."} Is this then the fate of any Lorentz-violating theory of gravity? Is one of the main motivations for considering emergent spacetime/LLI scenarios, understanding the thermodynamic aspects of gravity, also going to be its doom?

A potential breakthrough in this area was recently provided by the surprising realization that black holes in Lorentz-violating theories of gravity are quite different from the standard black holes in ordinary general relativity. 
Indeed, in such theories the Killing horizon does not capture the notion of the causal boundary in a black-hole spacetime, because these Lorentz-violating theories now admit superluminal excitations, which can cross the Killing horizon and escape to spatial infinity. It was recently found~\cite{Blas:2011ni, Barausse:2011pu, Saravani:2013kva} that the static, spherically-symmetric black-hole solutions in some specific Lorentz-violating theories contain a special hypersurface that acts as a genuine causal boundary because it traps \emph{all} excitations, even those which could be traveling at arbitrarily high velocities. 
This special hypersurface has been called the ``Universal horizon'' because no signal, moving at however high a velocity, can ever cross this surface to escape to infinity and is destined to hit the singularity. 

An interesting aspect of this finding, one that has direct relevance for the fate of the GSL, is that the Universal horizon is found to satisfy the first law of black-hole mechanics~\cite{Berglund:2012bu}. Not only that, the calculation of Ref.~\cite{Berglund:2012fk} to calculate the temperature, using the tunneling formalism of Ref.~\cite{Parikh:1999mf}, seems to predict the emission of a thermal flux from the Universal horizon.
But is this temperature associated with the Universal horizon relevant for observers outside the black hole? Or is the emitted radiation somehow reprocessed at the Killing horizon, in an energy and species dependent way? Is this the key to recovering a healthy thermodynamic behavior of black holes in Lorentz-violating theories? 
In order to answer these questions, as a preliminary step we need to understand particle dynamics on these spacetimes, and how it is affected by the presence of the Universal and Killing horizons.

The problem is that the natural ray trajectories to consider in these spacetimes correspond to particles with non-trivial dispersion relations, which adds complications as these trajectories are no longer metric geodesics, and are not determined by the spacetime geometry alone. How are the rays affected by the presence of Universal and Killing horizons?
Does the Universal horizon affect the ray trajectories of modified dispersion relations, in a way analogous to the Killing horizon affecting the relativistic rays? If so, what is the effect of the Killing horizon? Can we say anything about which surface is relevant for Hawking radiation? The purpose of this paper is to clarify these questions.

%Defining and understanding the surface gravity and temperature for universal horizons has already been the subject of some research~\cite{Berglund:2012bu, Berglund:2012fk, Cropp:2013zxi}. Understanding which surface radiates, and so should be relevant for thermodynamics, might help in restoring the Generalized Second Law, for which violations have been found in theories with multiple horizons~\cite{Dubovsky:2006vk, Eling:2007qd, Jacobson:2008yc}.

The paper is organized as follows: In Sec.~\ref{LV} we give a  brief introduction to Einstein-{\AEther} and \Horava--Lifshitz theories and the specific black-hole solutions that we study. We also discuss why the Universal horizon has not been seen earlier in the studies related to the causal structure of analogue black holes. 
In Sec.~\ref{physicaltrajectories} we review the behavior of relativistic rays at the Killing horizon and, as a warm-up, we perform a study of slices of constant ``\aether\ time''. We then proceed to the main body of our investigation in Sec.~\ref{physicaltrajectories} where we look at the ray trajectories associated to modified dispersion relations in these spacetimes. 
In addition to studying the behavior of rays near the Universal and Killing horizons, we also provide a notion of surface gravity for the Universal horizon, and compare it to those already existing in the literature~\cite{Berglund:2012bu, Berglund:2012fk,Cropp:2013zxi}. 
The issue of which surface is to be associated with thermal emission and the heuristic physical picture of Hawking radiation is discussed in Sec.~\ref{Hawkingrad}. We conclude in Sec.~\ref{discussion} with a summary and discussion of the implications of our work, and indicate some possible future directions. Appendix~\ref{gendispersion} discusses the universality of temperature of the Universal horizon. An alternative derivation of the conserved quantity and the particle trajectory  is given in appendix~\ref{apdx:conservation}.

%------------------------------------------------------------------------------------------------------------------------------------------
\section{Black Holes in Lorentz violating  theories}%------------------------------------------------------------------------------------------------------------------------------------------
\label{LV}
%------------------------------------------------------------------------------------------------------------------------------------------
Einstein--{\AEther} and \Horava--Lifshitz gravity are two well-studied theories of gravity that are diffeomorphism invariant but violate the local Lorentz invariance. Einstein--{\AEther} theory violates Lorentz invariance by introducing a preferred observer $u^a$, while \Horava--Lifshitz gravity introduces a preferred foliation defined by a scalar field $\tau$ called the Khronon. 

In this section, after giving a brief introduction to these two theories, we shall study the specific static spherically-symmetric black-hole solutions containing Universal horizons. We shall also explain why Universal horizons are not found in the acoustic black holes. 

%------------------------------------------------------------------------------------------------------------------------------------------
\subsection{Einstein--{\AEther} theory}
%------------------------------------------------------------------------------------------------------------------------------------------

Einstein--{\AE}ther theory, (for general background see Refs.~\cite{Jacobson:2000xp, Jacobson:2010mx, Gasperini:1987nq, Gasperini:1998eb, Jacobson:2008aj, Eling:2004dk}), is a Lorentz-violating theory which still maintains many of the nice features of General Relativity, such as general covariance and second-order field equations. This is done through introducing a timelike unit vector field, $u^a$, known as the {\ae}ther. The action is given by

\begin{equation}
S=\frac{1}{16\pi G}\int \d^4 x\sqrt{-g} (R+\mathcal{L}_{ae})\,; \qquad \mathcal{L}_{ae}=-Z^{ab}{}_{cd}\,(\nabla_au^c)(\nabla_b u^d)+\lambda(u^2+1).
\label{ac:ae}
\end{equation}
Here $\lambda$ is a Lagrange multiplier, enforcing the unit timelike constraint on $u^a$, and $Z^{ab}{}_{cd}$ couples the \aether\ to the metric through four distinct coupling constants:
\begin{equation}
Z^{ab}{}_{cd}=c_1g^{ab}g_{cd}+c_2\delta^a{}_c\delta^b{}_d+c_3\delta^a{}_d\delta^b{}_c-c_4 u^au^bg_{cd}.
\end{equation}
Throughout we will use a convenient shorthand to represent combinations of these constants;  such as $c_{14}=c_1+c_4$, \emph{etc}. 

%------------------------------------------------------------------------------------------------------------------------------------------
\subsection{\Horava--Lifshitz gravity}
%------------------------------------------------------------------------------------------------------------------------------------------

\Horava--Lifshitz (HL) gravity (see for example~\cite{Sotiriou:2010wn} for a review) was motivated by the possibility of achieving renormalizability by adding to the action the terms containing higher-order spatial derivatives of the metric, but no higher-order time derivatives, so as to preserve unitarity.   This procedure naturally leads to a foliation of spacetime into spacelike hypersurfaces.
%, labeled by the $t$ coordinate and with $x^i$ being the coordinates on each surface. The resulting theory is then invariant only under the reduced set of diffeomorphisms that leave this foliation intact, $t\to \tilde{t}(t)$ and $x^{i}\to \tilde{x}^{i}(t,x^i)$. 

Power-counting renormalizability requires the action to include terms with at least 6 spatial derivatives in 4 dimensions~\cite{Horava:2008ih,Horava:2009uw,Visser:2009fg}, but all lower-order operators compatible with the symmetry of the theory are expected to be generated by radiative corrections, so the most general action takes the form~\cite{Blas:2009qj}
\begin{equation}
\label{SBPSHfull}
S_{HL}= \frac{M_{\rm Pl}^{2}}{2}\int \d t\, \d^3x \, N\sqrt{h}\left(L_2+\frac{1}{M_\star^2}\;L_4+\frac{1}{M_\star^4}\;L_6\right)\,,
\end{equation}
where $h$ is the determinant of the induced metric $h_{ij}$ on the spacelike hypersurfaces, while
\begin{equation}
L_2=K_{ij}\,K^{ij} - \lambda K^2 
+ \xi\, {}^{(3)}\!R + \eta a_ia^i\,,
\end{equation}
with $K$ the trace of the extrinsic curvature $K_{ij}$, ${}^{(3)}\!R$ being the Ricci scalar of $h_{ij}$, $N$ the lapse function, and $a_i=\partial_i \ln N$. 
The quantities $L_4$ and $L_6$ denote a collection of 4th and 6th order operators respectively, and $M_\star$ is the scale that suppresses these operators (which does not coincide {\em a priori} with $M_{\rm Pl}$). 

%One interesting, and at the same time problematic, feature of HL gravity is the presence of a new propagating scalar mode associated with the reduced diffeomorphism invariance of the theory with respect to GR.  However, if one chooses to restore diffeomorphism invariance, then this mode manifests as a foliation-defining scalar field~\cite{Jacobson:2010mx}. This field allows also to make manifest the relation between the $L_2$  Lagrangian and the Einstein--Aether one.

In Ref.~\cite{Jacobson:2010mx} (see also Ref.~\cite{Afshordi:2009tt}) it was shown that  the solutions of  Einstein--\Aether\ theory are also the solutions of the infrared limit of \Horava--Lifshitz gravity if the \aether\ vector is assumed to be hypersurface orthogonal before the variation. More precisely, hypersurface orthogonality can be imposed through the local condition
\begin{equation}
u_\mu=\frac{\partial_\mu \tau} {\sqrt{-g^{\alpha\beta} \;\partial_\alpha \tau \, \partial_\beta \tau}}\, ,
\label{eq:orthogonal}
\end{equation}
where $\tau$ is a scalar field that defines a foliation (often named for this reason the ``khronon"). Choosing $\tau$ as the time coordinate one selects the preferred foliation of HL gravity, and the action (\ref{ac:ae}) reduces to the action of the infrared limit of \Horava--Lifshitz gravity, whose Lagrangian we denoted as $L_2$ in Eq.~(\ref{SBPSHfull}). 
The details of the equivalence of the equations of motions and the correspondence of the parameters of the two theories can be found in Refs.~\cite{Barausse:2012ny, Barausse:2012qh, Jacobson:2013xta}.

This fact is particularly relevant for the present investigation. Indeed we shall consider here static, spherically-symmetric black hole solutions in Einstein--{\AEther} for which the \aether\ field is always hypersurface orthogonal. Hence such solutions of Einstein--{\AEther} are also solutions of \Horava--Lifshitz gravity (at least in the infrared limit when one neglects the $L_4$ and $L_6$ contributions to the total Lagrangian). 
We shall therefore consider, from here on, black-hole solutions in Einstein--{\AEther} theory.

%------------------------------------------------------------------------------------------------------------------------------------------
\subsection{Einstein--{\AEther} black holes}
%------------------------------------------------------------------------------------------------------------------------------------------
\label{ae-bhs}
%------------------------------------------------------------------------------------------------------------------------------------------

Black holes in Einstein-{\AE}ther theory have been extensively considered in recent years (see for example Refs.~\cite{Barausse:2011pu, Blas:2011ni, Berglund:2012bu, Berglund:2012fk, Eling:2006ec, Mohd:2013zca}). 
Among the most striking results concerning these solutions was the realization --- in the (static and spherically-symmetric) black-hole solutions of both Einstein-{\AE}ther and \Horava--Lifshitz gravity ---  that they seem generically to be endowed a new structure that was soon christened the Universal horizon~\cite{Barausse:2011pu, Blas:2011ni}. 

These Universal horizons can be described as compact surfaces of constant khronon field (and radius) where the  khronon diverges (while nothing singular happens to the metric). 
Given that the khronon field defines an absolute time, any object crossing this surface from the interior would necessarily also move back in absolute time, (the {\ae}ther time), something forbidden by the definition of causality  in the theory. Another way of saying this is that even a particle capable of instantaneous propagation, (light cones opened up to an apex angle of a full 180 degrees, something in principle possible in Lorentz-violating theories), would just move around on this compact surface and hence be unable to escape to infinity. 
This explains the name of Universal horizon; even the superluminal particles would not be able to escape from the region it bounds.

\begin{figure}[!htb]
\centering
\includegraphics[scale=1.0]{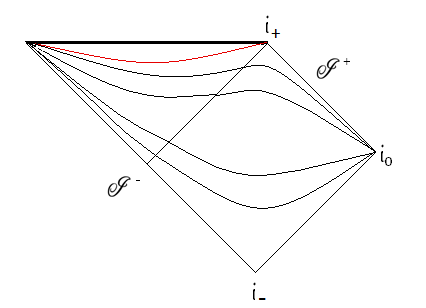}
\caption{Conformal diagram of black hole with Universal horizon, showing lines of constant khronon field, with the Universal horizon shown in red.}
\label{fig:conformal}
\end{figure}

The causal structure of these black-hole spacetimes is as shown in Fig.~\ref{fig:conformal}. The indicated hypersurfaces are constant khronon hypersurfaces. 
The special hypersurface behind the Killing horizon --- where the {\ae}ther and the Killing vector fields become orthogonal --- is the Universal horizon. Note that the Killing vector generates the time-translation isometry outside the Killing horizon, while inside it is spacelike. 

%------------------------------------------------------------------------------------------------------------------------------------------
%------------------------------------------------------------------------------------------------------------------------------------------

Coming back to black holes in Einstein--{\AEther} gravity, for some specific combinations of the coefficients there are explicit, exact solutions. In particular, two exact solutions for static, spherically symmetric black holes have been found. 
As we will use these solutions extensively throughout this paper, we will briefly summarize some of the relevant details of the solutions. For more information and background we refer the reader to Ref.~\cite{Berglund:2012bu}. Both solutions, in Eddington--Finkelstein coordinates, can be written as 
\begin{equation}
\d s^2 =-e(r)\;\d v^2 +2\,\d v\,\d r +r^2 \; \d \Omega^2.
\end{equation}
Here the form of the \aether\ is
\begin{equation}
u^a=\left\lbrace \alpha(r), \beta(r), 0, 0 \right\rbrace; \qquad u_a=\left\lbrace \beta(r)-e(r)\alpha(r), \alpha(r), 0, 0\right\rbrace.
\end{equation}
Note from the normalization condition, $u^2=-1$, there is a relation between $\alpha(r)$ and $\beta(r)$:
\begin{equation}
\beta(r)=\frac{e(r)\alpha(r)^2-1}{2\alpha(r)}.
\end{equation}
We can also define a spacelike vector $s^a$, such that
\begin{equation}
s^au_a=0; \qquad s^2=1.
\end{equation}
Explicitly 
\begin{equation}
s^a=  \left\{\alpha(r),e(r)\alpha(r)-\beta(r),0,0\right\} =   \left\lbrace \alpha(r), \frac{e(r)\alpha(r)^2+1}{2\alpha(r)}, 0, 0\right\rbrace,
% =\left\lbrace \alpha, \beta', 0, 0\right\rbrace, 
\label{eq:s}
\end{equation}
which clearly ensures $s^2=1$.
The two known exact black-hole solutions to Einstein--{\AEther} theory correspond to the special combinations of coefficients $c_{123} = 0$ and $c_{14}=0$.
%------------------------------------------------------------------------------------------------------------------------------------------
\begin{itemize}
%------------------------------------------------------------------------------------------------------------------------------------------
 \item Solution 1: $c_{123} = 0$.
%------------------------------------------------------------------------------------------------------------------------------------------

For this solution we have
\begin{equation}
e(r)=1-\frac{r_0}{r}-\frac{r_u(r_0+r_u)}{r^2}; \qquad\hbox{where}\qquad  r_u = \left[\sqrt{\frac{2 - c_{14}}{2(1 - c_{13})}} - 1\right]\frac{r_0}{2}.
\end{equation}
Here is $r_0$ is essentially the mass parameter. Furthermore
\begin{equation}
\alpha(r)=\left(1+\frac{r_u}{r}\right)^{-1}; \qquad \beta(r)=-\frac{r_0+2r_u}{2r}.
\end{equation}
It is also useful to decompose the Killing vector along $u$ and $s$ using the relations
\begin{equation}
\chi \cdot u=-1+\frac{r_0}{2r}\,; \qquad \chi\cdot s =\frac{r_0+2r_u}{2r}.
\end{equation}
For this particular exact solution, the Killing horizon is located at $r_{\KH}=r_0+r_u$, and the Universal horizon at $r_{\UH}={r_0}/{2}$.

%------------------------------------------------------------------------------------------------------------------------------------------
\item Solution 2: $c_{14} = 0$.
%------------------------------------------------------------------------------------------------------------------------------------------

For this solution we have
\begin{equation}
e(r) = 1 - \frac{r_0}{r} - \frac{c_{13}r_{\ae}^4}{r^4}; \qquad r_{\ae} = \frac{r_0}{4}\left[\frac{27}{1-c_{13}}\right]^{1/4};
\end{equation}
\begin{equation}
\alpha(r) = \frac{1}{e(r)}\left(-\frac{r_{\ae}^2}{r^2} +\sqrt{e(r) + \frac{r_{\ae}^4}{r^4}}\right); \qquad \beta(r) = -\frac{r_{\ae}^2}{r^2}~.
\end{equation}

Furthermore, the Killing vector is decomposed as 
\begin{equation}
\chi \cdot u =-\sqrt{1-\frac{r_0}{r}+\frac{(1-c_{13})r_{\ae}^4}{r^4}}\, ; \qquad \chi \cdot s =\frac{r_{\ae}^2}{r^2}~.
\end{equation}
The Killing horizon is located by solving the quartic polynomial $e(r)=0$, so
\begin{equation}
r_{\KH}=\frac{r_0}{4}+\frac{A}{8}+\frac{1}{8}\sqrt{8r_0^2-{\frac {6c_{13}r_0^2}{{\zeta}^{2}}}+{\frac {6r_0^2\zeta^2}{1-c_{13}}}+\frac{16 r_0^3}{A}},
\end{equation}
where
\begin{equation}
A = r_0 \sqrt{ 4 + {6c_{13}\over\zeta^2} + {6\zeta^2\over (c_{13}-1)}}; \qquad  \zeta=  \sqrt[6]{c_{13} \left((c_{13}-1) ^{2}+\sqrt { \left( 1-c_{13} \right) ^{3}} \right)}\, . 
\end{equation}
The Universal horizon is located at $r_{\UH}=3\, r_0/4$.

%------------------------------------------------------------------------------------------------------------------------------------------
\end{itemize}
%------------------------------------------------------------------------------------------------------------------------------------------

%------------------------------------------------------------------------------------------------------------------------------------------
\subsection{A digression on acoustic gravity and Universal horizons}
%------------------------------------------------------------------------------------------------------------------------------------------
\label{digression}
%------------------------------------------------------------------------------------------------------------------------------------------

It is interesting to note that in the menagerie of acoustic spacetimes, a specific causal structure similar to that of these Einstein-\AEther\ black holes was found (see Fig.~26 of Ref.~\cite{Barcelo:2004wz}). 
This spacetime was named the ``unphysical black hole" as it needed an unphysical, diverging, fluid flow in order to mimic the presence of a singularity. Nonetheless, the resemblance with the causal structure shown in Fig.~\ref{fig:conformal} appears striking.
However, no concept of Universal horizon has ever surfaced in acoustic spacetimes (or any of the other analogue models for that matter).  It is therefore interesting to clarify why this is the case. 

Of course, it is trivial to put the metric of the two previously described black-hole solutions of Einstein--{\AE}ther theory into the standard Painleve--Gullstrand form,
\begin{equation}
\d s^2=-\left(1-\frac{ {\rm v}^2 }{c_s^2}\right)\d t^2 +2\,\frac{{\rm v}}{c_s}\,\d t\, \d r + \left|\d \vec{x}\right|^2,
\end{equation}
where ${\rm v}$ is the flow velocity and $c_s$ the speed of sound for the corresponding acoustic geometry. In this geometry there is a natural notion of the preferred frame, the frame in which the fluid is at rest. So the corresponding {\ae}ther field is most naturally taken to be
\begin{equation}
u^a=\left\lbrace u^t, \frac{\rm v}{c_s}\right\rbrace.
\end{equation}
The zeroth component is determined by the requirement that the {\ae}ther field is unit timelike, which gives
\begin{equation}
u^t=\frac{{\rm v}^{2}\pm\sqrt{ {\rm v}^4-c_s^4+{\rm v}^2\,c_s^2} }{c_s^2-{{\rm v}^2}}.
\end{equation}
Regularity at the Killing horizon (the surface where ${\rm v}=c_s$) fixes the sign to be minus. Now let us calculate --- when is the Killing vector orthogonal to the {\ae}ther?  We have 
\begin{equation}
\chi \cdot u = \pm \frac{ \sqrt{ {\rm v}^{4}+c_s^4-{\rm v}^2 c_s^2} } {c_s^2} 
=\pm \frac{ \sqrt{ ({\rm v}^{2}-c_s^2)^2 + ({\rm v} c_s)^2} } {c_s^2} .
\end{equation}
But this, being proportional to a sum of squares, is never zero for real-valued $\rm{v}$. Therefore, there is no Universal horizon. 

Why is there no Universal horizon in this case? In the acoustic geometry, the \aether\ field is determined by the same flow, ${\rm v}$, that determines the metric. 
There is simply not enough freedom to have additional structures such as Universal horizon, where the geometry is well behaved, but the \aether\ time has run to infinity. One needs a separately defined dynamical {\ae}ther to have a Universal horizon. 
This of course does not mean that we cannot capture this structure within the context of analogue models. One simply needs more degrees of freedom in order to do that. 
For example, one idea would be to consider a charged fluid in an external magnetic field. This would introduce the velocity of the charge flow which is different from the velocity of energy flow (the fluid velocity). It would be interesting to construct such a model in detail to see if one can simulate the Universal horizon within the context of analogue models.

%------------------------------------------------------------------------------------------------------------------------------------------
\section{Physical trajectories in an Einstein--\Aether\ black hole}
%------------------------------------------------------------------------------------------------------------------------------------------
\label{physicaltrajectories}
%------------------------------------------------------------------------------------------------------------------------------------------

We now turn to analyzing the motion (determined by the group velocity) of particles endowed with modified, Lorentz-violating, dispersion relations in the black-hole geometries discussed in Sec.~\ref{ae-bhs}. 
Modified dispersion relations arise due to the interaction of these particles with the {\ae}ther. Thus the trajectories are \emph{not} simply the geodesics determined from the spacetime metric.  

It will prove beneficial to first review the standard description of particle trajectories in a black-hole spacetime, with particular attention to their behavior close to the Killing horizon. 
We do that in   Secs.~\ref{sec:standardpeeling} and ~\ref{sec:RaeT}. After discussing the modified dispersion relations in Sec.~\ref{subsec:modDisRel}, we we will introduce an appropiate notion of the conserved particle energy in Sec.~\ref{subsec:conservation law}, which will then be needed in Sec.~\ref{subsec:trajectories} where we finally construct the ray trajectories.

%------------------------------------------------------------------------------------------------------------------------------------------
\subsection{Ray tracing and peeling in purely metric black holes}
%------------------------------------------------------------------------------------------------------------------------------------------
\label{sec:standardpeeling}
%------------------------------------------------------------------------------------------------------------------------------------------

Let us then first briefly recap some aspects of ray tracing, taking as a simplest example the Schwarzschild spacetime. (See also~\cite{Poisson}.) In Eddington--Finkelstein coordinates
\begin{equation}
\d s^2= -\left(1-\frac{2M}{r}\right)\d v^2+2\,\d v\,\d r+ r^2\d \Omega_2^2,
\end{equation}
the outgoing rays will be given by 
\begin{equation}
\left.\frac{\d v}{\d r}\right|_{\mathrm{out}}=\frac{2}{1-\frac{r}{2M}}
\end{equation}
with the ingoing rays travelling at $45^{\circ}$. 

\begin{figure}[!htb]
\centering
\includegraphics[scale=0.5]{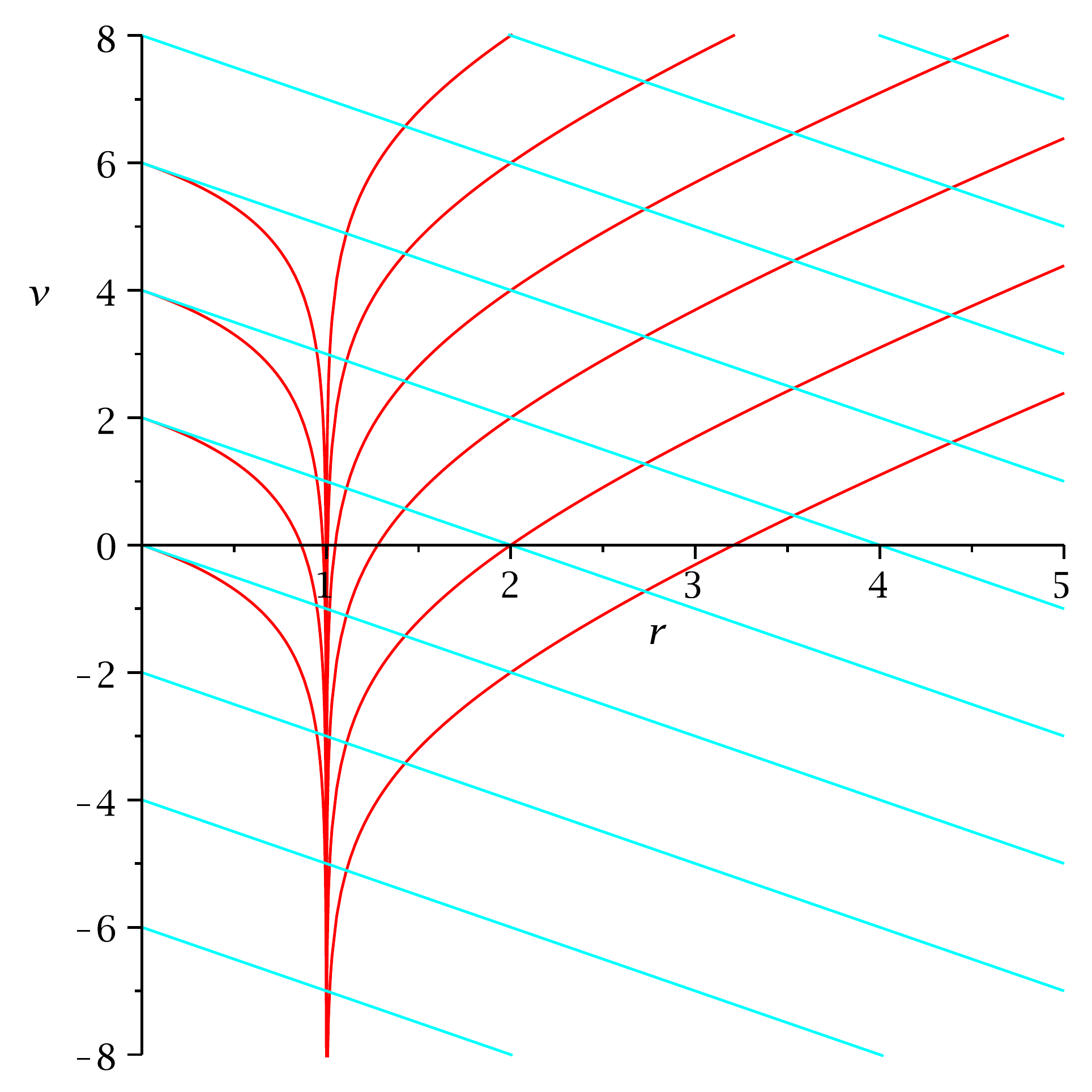}
\caption{Ingoing (blue) and outgoing null rays (red) for standard Schwarzschild black hole, horizon at r=1.}
\label{schwarz}
\end{figure}

In these coordinates a peeling-off of rays from the Killing horizon can be seen (Fig.~\ref{schwarz}). 
It is this peeling that is related to the large increase in frequency/energy as one traces back along an outgoing mode in Hawking radiation~\cite{Jacobson:1999zk}.

%----------------------------------------------------------------
%----------------------------------------------------------------

In particular one can associate a surface gravity notion to this peeling by a suitable expansion of the ray behaviour close to the Killing horizon. This takes the form

\begin{equation}
\left.\frac{\d r}{\d v}\right|_{\mathrm{out}}=\left.\frac{\d r}{\d v}\right|_{\KH}+\left.\frac{\d}{\d r} \frac{\d r}{\d v}\right|_{\KH}(r-r_{\KH}) +\mathcal{O}(r-r_{\KH})^2.
\label{eq:standarddrdv}
\end{equation}
At the Killing horizon, the first term on the right hand side of this equation vanishes, and the second term defines the peeling surface gravity
\begin{equation}
\kappa_{\rm peeling}\equiv \left. \frac{1}{2}\frac{\d}{\d r} \frac{\d r}{\d v}\right|_{\KH}
\end{equation}

There are other ways to define surface gravity, all of which coincide for stationary black holes in general relativity, but this particular version is that most closely linked to the trajectories of particles, and therefore of most utility in a ray tracing study. 
Furthermore, this version of surface gravity  is closely linked to Hawking radiation when the degeneracy between definitions is broken (see~\cite{Barcelo:2010xk, Cropp:2013zxi}).

%------------------------------------------------------------------------------------------------------------------------------------------

Strengthened by this brief review of the standard case, we can now consider the propagation of rays in the presence of the {\ae}ther.  
However, before doing so, we find it useful to perform a study of the slices of constant khronon field $\tau$ in our spacetime, as these are closely related to the Universal horizon and also can be seen as the ray trajectories of physical particles in the limiting case of an infinite propagation speed (with respect to the aether).

%------------------------------------------------------------------------------------------------------------------------------------------
\subsection{Rays of constant \aether\ time}
%------------------------------------------------------------------------------------------------------------------------------------------
\label{sec:RaeT}
%------------------------------------------------------------------------------------------------------------------------------------------

\begin{figure}[!h]
\centering
\includegraphics[scale=0.5]{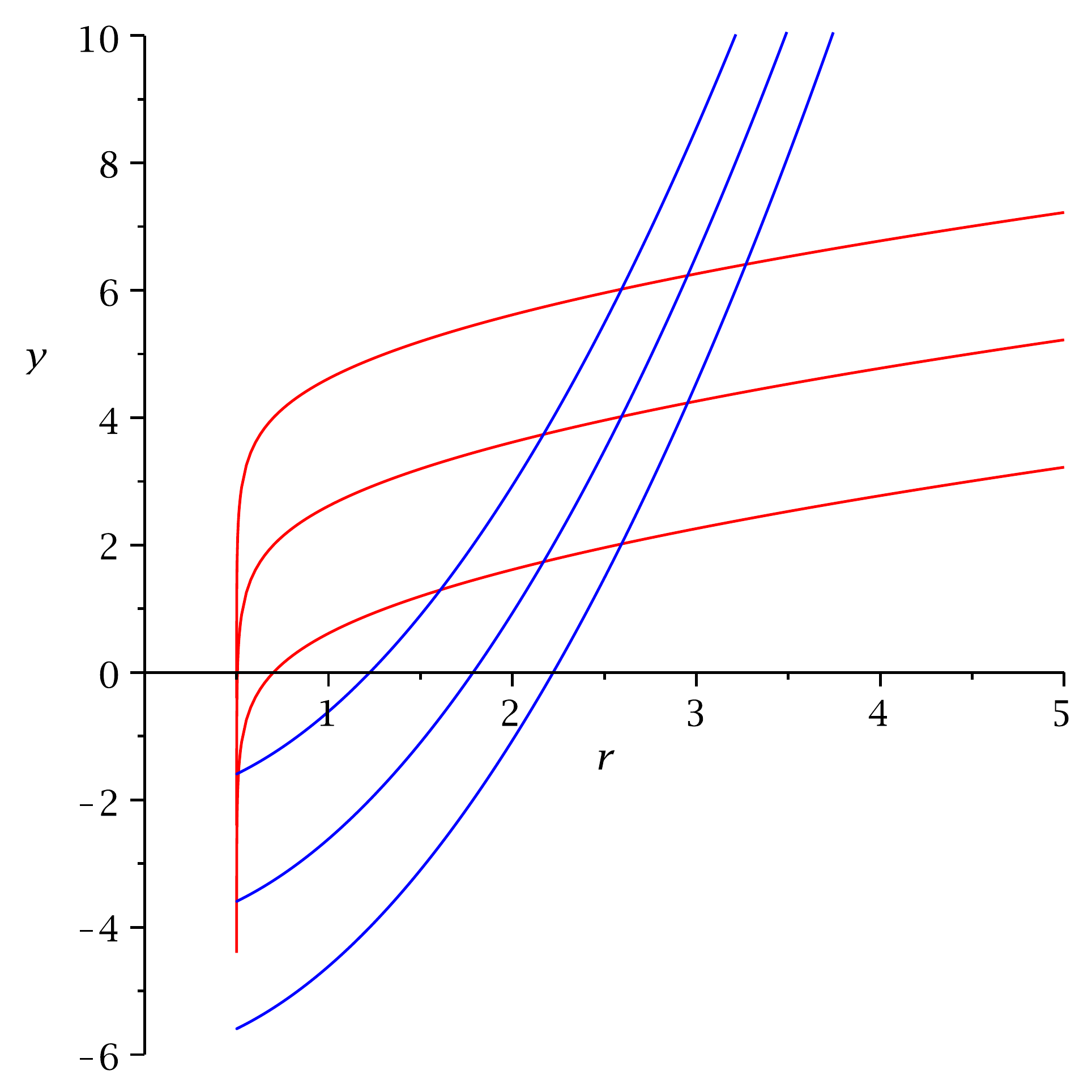}
\caption{Lines of constant $\tau$ (red) and $\sigma$ (blue) for the $c_{123}=0$ solution. The Killing horizon is at $r=1$ and the Universal horizon at $r=1/2$.}
\label{fig:tau-sigma}
\end{figure}

As noted, we have a clear notion of causality in these theories: Nothing can travel backwards in \aether\ time. 
The \aether\ time can be related to the metric one (what we might call the Killing time, given that we are dealing with static metrics) given that the \aether\ one-form is
\begin{equation}
u=u_v\d v+u_r\d r= u_v\left(\d v +\frac{u_r}{u_v} \,\d r\right)= u_v\d \left( v +\int\frac{u_r}{u_v}\,\d r \right).
\end{equation}
Now, using the fact that $u^a$ has a unit norm, it can verified that there exists a function $\tau $ such that $u=\d \tau/||\d \tau||$. The explicit form of $\tau$ is given by
\begin{equation}
\tau =v+\int \frac{u_r}{u_v} \;\d r.
\end{equation}
Likewise, we can define a spatial $\sigma$ coordinate, corresponding to constant $s$ slices, $s=\d \sigma/||\d \sigma||$, and in a similar manner, find
\begin{equation}
\sigma =-v+\int \frac{s_r}{s_v} \;\d r.
\end{equation}
%Note that, at infinity the lines of constant $\tau$, $\sigma$ will agree with %lines of constant $v$, $r$, since at spatial infinity the \aether\ vector and %the Killing vector are aligned. 
We plot the constant $\tau$,~$\sigma$ surfaces in Fig.~\ref{fig:tau-sigma} for the $c_{123}=0$ solution.
%------------------------------------------------------------------------------------------------------------------------------------------
%------------------------------------------------------------------------------------------------------------------------------------------
A peeling-like behavior at the Universal horizon is evident, and indeed a notion of $\kappa_\mathrm{peeling}$ can be associated to the constant $\tau$ slices.  
As realistic rays must travel forward in \aether\ time they must have paths as steep as, or steeper than, the constant $\tau$ slices. Thus the constant $\tau$ slices, corresponding to infinite velocity with respect to the \aether\, will provide a lower bound to the peeling properties (and so the peeling surface gravity) of physical rays. 

We can easily calculate the value of this surface gravity which, as noted, will be the relevant one for rays propagating with infinite group velocity $\rm v_g$. 
Generically any particle propagating in our spacetimes will have a four-velocity that can be given in the orthonormal frame provided by $u^a$ and $s^a$ as,
\begin{equation}
V^a=u^a+{\rm v}_g\, s^a . 
\label{eq:fourV}
\end{equation}
The trajectory for an instantaneously propagating ray would then be given by
\begin{equation}
\frac{\d v}{\d r} = \frac{V^v}{V^r}= \lim_{{\rm v}_g \to \infty} \,\frac{u^v+{\rm v}_g s^v}{u^r+{\rm v}_g s^r}= \frac{s^v}{s^r}.
\label{eq:dvdr}
\end{equation}
Then, Taylor expanding the trajectory close to the horizon one gets
\begin{align}
\left.\frac{\d r}{\d v}\right|_{\mathrm{out}}=\left.\frac{\d r}{\d v}\right|_{\UH}+\left.\frac{\d}{\d r} \frac{\d r}{\d v}\right|_{\UH}(r-r_{\UH}) +\mathcal{O}(r-r_{\UH})^2.
\end{align}
A straightforward calculation based on Eq.~\eqref{eq:s} shows that $({\d r}/{\d v})|_{\UH}=0$ for both the aforementioned solutions. (This happens because $s^r|_{UH}=0$, which can be checked by plugging in either of the explicit formulae given in Sec.~\ref{ae-bhs}). 
This is also what one should expect from the behavior in Fig.~\ref{fig:tau-sigma}. Consequently,
\begin{align}
\kappa_{\UH}\equiv \left.\frac{1}{2}\frac{\d}{\d r} \frac{\d r}{\d v}\right|_{\UH}
\label{eq:kappapeelUH}
\end{align}
where $\kappa_{\UH}$ is by definition the ``surface gravity" corresponding to the peeling-off property of the infinite velocity modes close to the horizon. 
%This would be the surface gravity, which could have an associated temperature, of an infinitely fast mode, or any mode that has a  divergent velocity at the universal horizon. 

%------------------------------------------------------------------------------------------------------------------------------------------
%\subsection{Physical ray trajectories}
%------------------------------------------------------------------------------------------------------------------------------------------
%This introduces a conceptual problem as, from the geometric point of view, these particles are no longer free, they explicitly couple to the \aether\ field. \emph{Therefore particles no longer follow spacetime geodesics.}
%We can tackle this problem by using the fact that we still have a static spacetime, and therefore conservation of some notion of energy. Together with a specific choice of modified dispersion relation this will be seen to be enough to solve for the trajectory of a ray.
%

%%------------------------------------------------------------------------------------------------------------------------------------------
\subsection{Modified dispersion relations}
%------------------------------------------------------------------------------------------------------------------------------------------
\label{subsec:modDisRel}
%%------------------------------------------------------------------------------------------------------------------------------------------

In a Lorentz-violating scenario, particles will be generically coupled to the \aether\ and such a coupling will imply modified dispersion relations which can be naturally assigned in the preferred, \aether\ frame. Note that owing to the dynamical nature of {\ae}ther, modified dispersion relations do not cause any inconsistency in the Bianchi identities as discussed in Ref.~\cite{Kostelecky:2003fs}.
Let us stress that Lorentz-violation in the gravitational sector is expected to percolate into the matter sector via radiative corrections, at least in an effective field theory framework.  
This implies that even starting with a  Lorentz invariant matter sector our theory will end up  providing modified dispersion relations for all particles. Given that the gravitational theories under consideration are invariant under parity, we shall assume that no parity violating operators are induced in the matter sector (and that none of them are present at the tree level). 
For radial motion one can then generically expect modified dispersion relations of the form
\begin{equation}
\omega^2=c^2\, \ks^2+\ell^2\, \ks^4+\ell^4\, \ks^6 + \dots
\label{eq:mod-disp}
\end{equation}
%where frequency and wavenumber are to be taken to be those measured by an observer co-moving with the \aether\ $u^a$.
where $\omega\equiv-(k \cdot u)$ is the energy in the \aether\ frame (note that $\omega > 0$ for propagating particles) and (for radial motion) $\ks\equiv (k \cdot s)$ is the spatial component of $k_a$ orthogonal to the \aether\ field. 
The length scale $\ell$ here is the UV Lorentz-violating scale for matter which we do not need to necessarily identify with the Planck scale. Letting $\ell\to0$ one would recover standard Lorentz-invariant dispersion relations. Modified dispersion relations in the context of Einstein-{\AE}ther theory are also discussed in Ref.~\cite{Jishnu_thesis}.\par
Note that we are considering only UV modifications of the matter dispersion relations, while of course \emph{a priori} one might also expect radiative corrections to produce IR modifications (for e.g., inducing particle-dependent coefficients of the order $\ks^2$ terms), see for e.g. Ref.~\cite{Collins:2004bp}.
We do not explicitly consider these potential complications as they are highly constrained phenomenologically,  and furthermore there are known mechanisms that suppress such terms (see, for e.g., Ref.~\cite{Liberati:2013xla} and references therein). 
Also, in general a modified speed of light will not effect our results except for the fact that different particles will perceive different Killing horizons (placed at different radii). One might still wonder as to why the coefficients of $k_s$ appearing in the dispersion relation are not position dependent. The reason is simply that after going to an orthonormal frame these coefficients can depend only on a very limited number of scalars, built from gradients of $k_s$, extrinsic curvature of the {\ae}ther flow, and curvature.  The eikonal approximation kills the gradients of $k_s$, while those geometric scalars which could possibly occur are all of the curvature scale, and hence are suppressed by the curvature radius, which is assumed to be very large. There are also possible foliation dependent terms like extrinsic curvature. But extrinsic curvature is just the gradient of {\ae}ther, which determines the energy-momentum of the æther. This in turn is related to the curvature scale by the equation of motion. Thus consistency requires that such terms are also suppressed by the curvature radius. Therefore, in the ray approximation, for a large mass black hole, the dispersion relation does not have any spacetime dependent geometric coefficients. \par
Hence, we shall only consider $\omega^2=c^2 \,\ks^2+\ell^2 \,\ks^4$, which we shall further brutally simplify by looking at the expansion for low $k$
\begin{equation}
\label{eq:cubicDisRel}
 \omega=c \,\ks+\frac{1}{2c}\ell^2\, \ks^3.
\end{equation}
We do this mainly for simplicity, (in particular, by solving the cubic, $\ks(\omega)$ can be written down in a closed form).  Henceforth we will absorb the factor of $2c$ into the suppression factor.  Later on we shall consider the possibility of more general dispersion relations and show that our main results are independent of the specific form of the dispersion relation. 

%------------------------------------------------------------------------------------------------------------------------------------------
\subsection{A notion of conserved energy}
%------------------------------------------------------------------------------------------------------------------------------------------
\label{subsec:conservation law}
%------------------------------------------------------------------------------------------------------------------------------------------

 The modified dispersion relation gives us \emph{one} equation between $\omega$ and  $\ks$. In order to find their explicit values on spacetime we need a \emph{second} equation relating them. This equation is a conservation equation, which says that $k \cdot \chi$ is constant on the whole spacetime. We now give the argument for general modified dispersion relations. Another derivation is given in  appendix~\ref{apdx:conservation}. \par
Firstly, define differential operators acting on a generic quantity $X$ by setting
\begin{equation}
\nabla_1 X = u^a \partial_a X, \qquad\hbox{and} \qquad \nabla_2 X = \sqrt{\nabla_a[(g^{ab}+u^au^b)\nabla_b X]}. 
\end{equation}
These are respectively temporal derivatives in the direction of the \aether, and spatial derivatives on constant \aether-time hypersurfaces.  
Then to a given dispersion relation, $\omega=f(\ks)$, we can naturally associate the differential operator
\begin{equation}
D(x;\nabla) =   -(\nabla_1)^\dagger (\nabla_1)+ [f(-i\nabla_2)]^\dagger [f(-i\nabla_2)]\, .
\end{equation}
%(by associate we mean that the above operator would provide the dispersion relation once e.g. one applies a plane wave to it).
We are working with a static spacetime, and therefore the Killing equation implies that in terms of the Killing time coordinate
\begin{equation}
[-i\partial_t, D(x,\nabla) ] = 0.
\end{equation}
Both of these are Hermitian operators, so the vanishing of their commutator implies simultaneous diagonalizability.
This implies that $-i\partial_t$ is explicitly diagonalizable, so one may write
\begin{equation}
\Phi(t,r) = e^{i\Omega t}\; \Phi(r),
\end{equation}
with $\Omega$ a \emph{position independent} constant. 
%Note that 
%\begin{equation}
%D(x;\nabla)  [e^{i\Omega t} \; \Phi(x)] = e^{-\Omega t} \; D(r;i\Omega\nabla_i) \,\Phi(r).
%\end{equation}
%{\bf Warning: the above formula needs to be double checked! Do we need it?}\\

Now pick a particular tetrad based on completing the zwei-bein $(u^a,\,s^a)$. Using spherical symmetry one can decompose the 4-momentum as
\begin{equation}
k_a = e^A{}_a k_A = f(\ks) \;u_a + \ks \;s_a\, ,
\end{equation}
from which we can read
\begin{equation}
- \Omega = k_t =  f(\ks) \;u_t + \ks \; s_t, 
\end{equation}
which implicitly defines $\ks(r)$ via
\begin{equation}
-\Omega =  f(\ks(r)) \;u_t(r) + \ks(r) \;s_t(r).
\end{equation}
%Once we have $\ks(r)$, note that
%\begin{equation}
%k_r(r) =   f(\ks(r)) \;u_r(r) + \ks(r) \;s_r(r),
%\end{equation}
%and so the phase is
%\begin{equation}
%\Theta = \int k_a \d x^a = \Omega t + \int k_r \; \d r = \Omega t + \int \{   f(\ks(r)) \;u_r(r) + \ks(r) \;s_r(r) \} d r.
%\end{equation}
In particular, we now have the statement
\begin{equation}
\nabla_a \left(k_b\, \chi^b\right) = \nabla_a \Omega = 0.
\end{equation}
The fact that we have this position-independent constant means we can solve for $\ks(r)$ and $\omega(r)$. Then we can integrate the group velocity to obtain the ray trajectory of the particle explicitly. 
Physically, $\Omega$ is the Killing energy at infinity, where also $\Omega=\omega$. That is, when we talk about high (low) $\Omega$ rays, we mean rays that arrive at infinity with high (low) Killing energy.

%------------------------------------------------------------------------------------------------------------------------------------------
\subsection{Physical ray trajectories}
%------------------------------------------------------------------------------------------------------------------------------------------
\label{subsec:trajectories}
%------------------------------------------------------------------------------------------------------------------------------------------
We now are set to solve for the trajectory, with our cubic dispersion relation in Eq.~\eqref{eq:cubicDisRel}
\begin{equation}
\omega=\ks+\ell^2\ks^3,
\end{equation}
and the conservation equation,
\begin{equation}
\label{eq:conservation eqn}
-\Omega =  \omega \,(\chi \cdot u) \pm \ks\, (\chi \cdot s).
\end{equation}
The $\pm$ refers to whether the mode is outgoing or ingoing, respectively. 
In what follows we will focus on the outgoing modes.
As $\Omega$ is the energy in the \aether\ frame at spatial infinity,  we require that $\Omega$ be non-negative. \par
What happens to $\ks$ and $\omega$ near the Universal horizon? 
On the Universal horizon we have $\chi \cdot u = 0$ and $\chi \cdot s = \|\chi \|$. We might naively but incorrectly argue from Eq.~\eqref{eq:conservation eqn} that $ - \Omega = \ks \|\chi \|$, but this is inconsistent because the RHS is positive while the LHS is negative. 
This just points out the fact that something singular is happening on the Universal horizon. Let us then seek to better understand the divergence structure of $\ks$ on the Universal horizon.

Note  that $u$ is everywhere timelike while $\chi$ is timelike outside, null on, and spacelike inside the Killing horizon. 
Also from the fact that $(\chi\cdot u)=-1$ at infinity and becomes zero only at the Universal horizon, we deduce that this product is  negative everywhere outside the Universal horizon, we hence write it as $-|\chi \cdot u|$. 
For the sign of $\chi \cdot s$ one has to specify a choice of the $s$ basis (inward or outward pointing). We choose $s$ in Eq.~\eqref{eq:s} so that  $\chi \cdot s$  is  positive everywhere  outside the Universal horizon. 
We hence write it as $| \chi \cdot s |$. So we may rewrite Eq.~\eqref{eq:conservation eqn} as
 \begin{equation}
\label{eq:conservation eqn2}
 \Omega = (\ks+ \ell^2 \ks^3)\, |\chi \cdot u | - \ks\, |\chi \cdot s|.
 \end{equation}
Let us parametrize the singular behavior of $\ks$ at the Universal horizon as
\begin{equation}
 \ks \sim \frac{a(r)}{| \chi \cdot u |^\gamma},
 \end{equation}
where $a(r)$ is regular at the Universal horizon and $\gamma$ is a constant to be determined. 
Substituting this ansatz in Eq.~\eqref{eq:conservation eqn2}, finiteness of the LHS implies 
\begin{equation}
\gamma=\frac{1}{2}\,; \qquad \hbox{and} \qquad a(r_{\UH})=\dfrac{\sqrt{ |\chi \cdot s|}_{\UH}}{\ell}.
\end{equation}
Therefore, close to the Universal horizon $\ks \sim 1/\sqrt{ |\chi \cdot u|}$.

%------------------------------------------------------------------------------------------------------------------------------------------
\begin{figure}[!htb]
\centering
\includegraphics[scale=0.8]{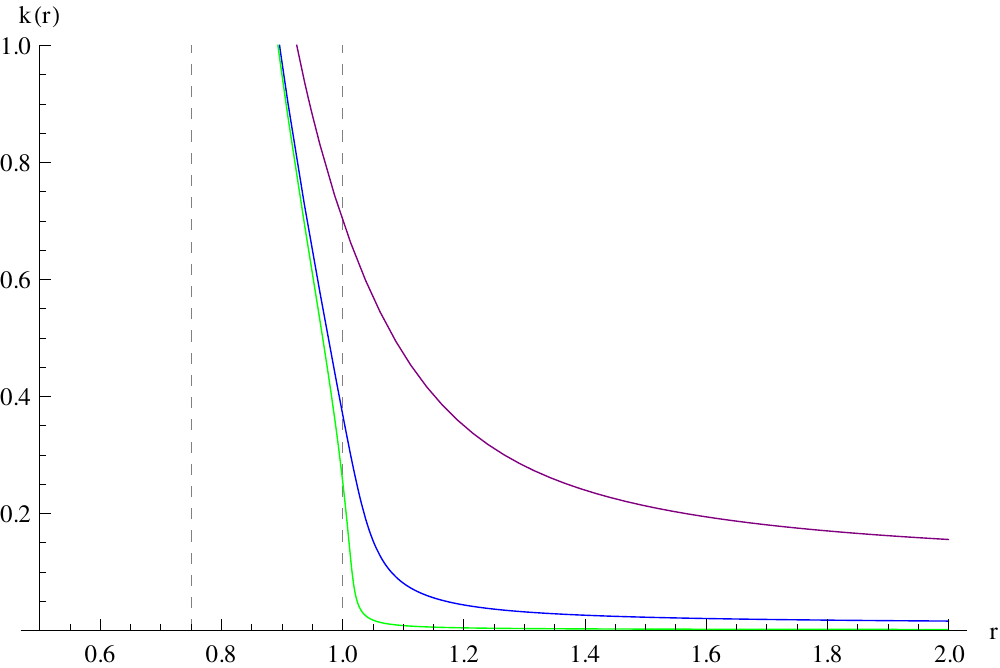}
\includegraphics[scale=0.8]{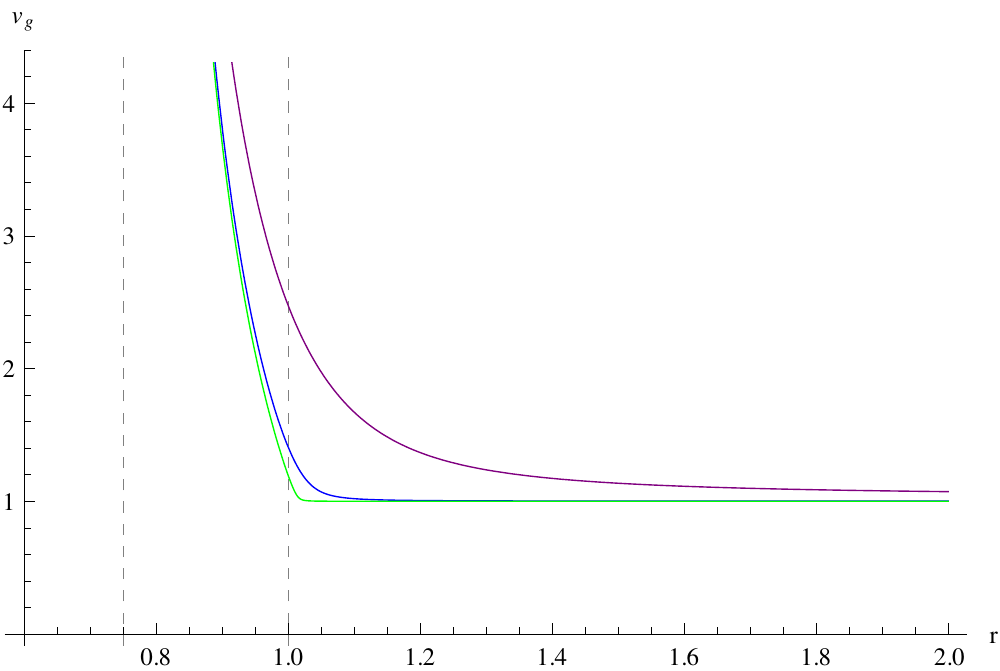}
\caption{The wavenumber $k$ (left) and group velocity (right) as functions of $r$  for the $c_{14}=0$ solution, for energies $\Omega = 10^{-1}$ (purple), $\Omega = 10^{-2}$ (blue) and $\Omega = 10^{-3}$ (green). The Lorentz-violating scale is fixed at $\ell=1$. The Killing horizon is at $r=1$, and the Universal horizon is at $r=0.75$. }
\label{fig:k}
\end{figure}
%------------------------------------------------------------------------------------------------------------------------------------------

Let us now move away from the Universal horizon and consider the trajectory throughout the whole spacetime. For doing this we shall basically follow the procedure of section \ref{sec:RaeT}, with the difference that the group velocity $\partial \omega/\partial \ks$ is no longer infinite, but is now a function of $r$ which diverges at the Universal horizon. 
From our dispersion relation in Eq.~\ref{eq:cubicDisRel} we find the group velocity to write the four velocity of the particle as in Eq.~\eqref{eq:fourV}, $V^a = u^a + {\rm{v}}_g \; s^a$.  
In the Eddington-Finkelstein coordinate system denoted by $\{v,r\}$, the trajectory of the particle can again be obtained by solving the differential equation
\begin{equation}
\label{eq:trajectory}
\frac{\d v}{\d r}=\frac{V^v}{V^r}=\frac{u^v (r)+ {\rm v}_g(\ks)\; s^v(r)}{u^r (r)+ {\rm v}_g(\ks) \; s^r(r)}.
\end{equation}
When one aims at plotting the above trajectories in spacetime it is clear that the only difficulty arises from the $\ks$-dependence of ${\rm v}_g=1+3 \ell^2\, \ks^2$, as the 3-momentum $\ks$ is \emph{not} a conserved quantity along the path. 
This is where the conservation equation, Eq.~\eqref{eq:conservation eqn2}, comes in as it allows us to solve for $\ks$ as a function of $r$ and $\omega$,
\begin{equation}
\ks(r)=\frac{\left[(12)^{\frac{1}{2}}\left( 9\Omega\ell +\sqrt{\frac{\textstyle 12(\chi\cdot s-\chi\cdot u)^3}{\textstyle(\chi\cdot u)}+(9\Omega\ell)^2}\right)(\chi\cdot u)^2\right]^{\frac{2}{3}}+(12)^{\frac{2}{3}}(\chi\cdot u)(\chi\cdot u-\chi\cdot s)}{6(\chi\cdot u) \ell \left[ \left( 9\Omega\ell +\sqrt{\frac{\textstyle 12(\chi\cdot s-\chi\cdot u)^3}{\textstyle (\chi\cdot u)}+(9\Omega\ell)^2}\right)(\chi\cdot u)^2\right]^{1/3}},
\end{equation}
where the functions $\chi\cdot u$ and $\chi \cdot s$ can be read off for either of the two particular solutions given in section \ref{ae-bhs}. 
(The particular form of $\ks(r)$ given above of course depends crucially on the assumed cubic dispersion relation. For more general dispersion relations, while $\ks(r)$ certainly exists, it may be difficult to exhibit an explicit formula.)

\begin{figure}[!htb]
\centering
\includegraphics[scale=0.75]{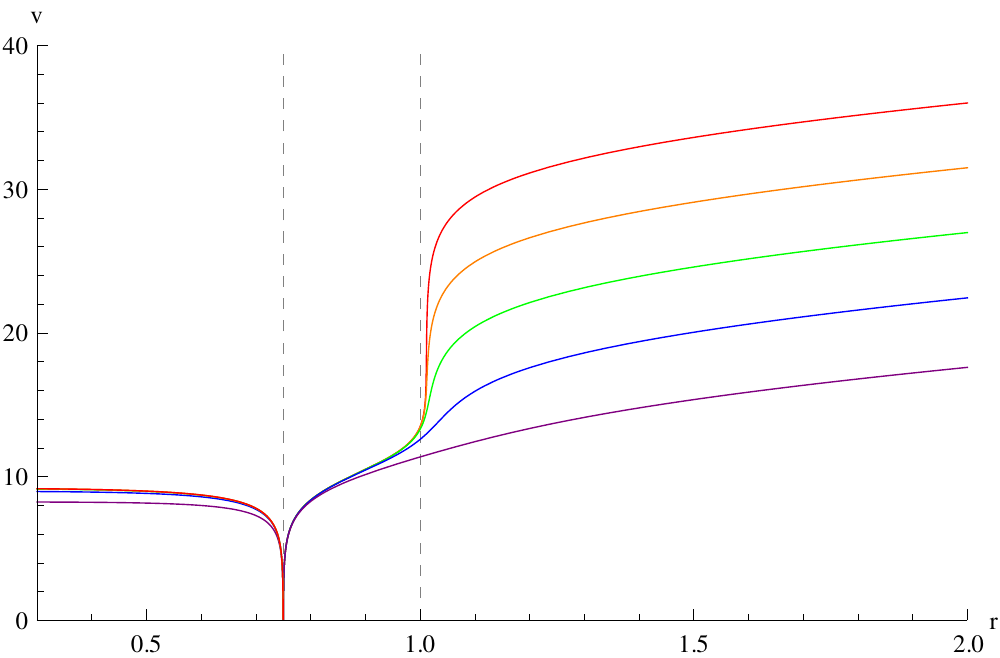}
\includegraphics[scale=0.75]{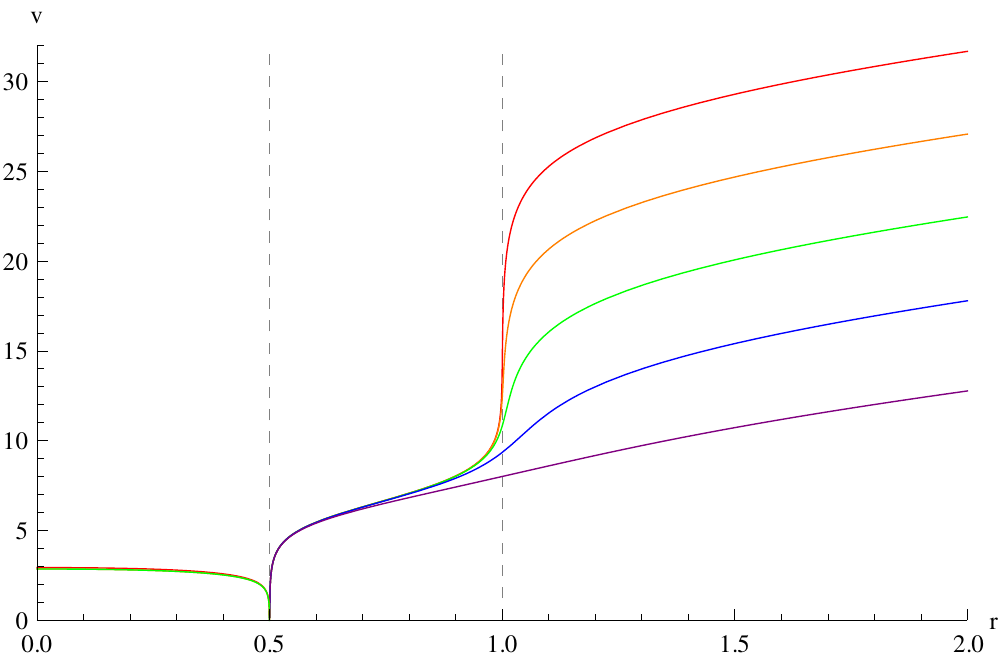}
\caption{Trajectories of the outgoing particle in $v$--$r$ Eddington--Finkelstein coordinates. \newline Energies of $\Omega = 0.1$ (purple), $\Omega=10^{-2}$ (blue), $\Omega=10^{-3}$ (green), $\Omega=10^{-4}$ (orange) and $\Omega=10^{-5}$ (red). For these parameters of the black hole the  $c_{123}=0$ solution (left) has Universal horizon at $r_{\UH}=0.75$, while for the $c_{14}=0$ solution (right)  the Universal horizon is at $r_{\UH}=0.5$. For both situations $r_{\KH}=1$. Behaviour at the Universal horizon is universal while behaviour at the Killing horizon at $r_{\KH}=1$ depends on energy.}
\label{fig:lingering}
\end{figure}

We can now numerically integrate Eq.~\eqref{eq:trajectory} and plot  trajectories for different values of the conserved energy $\Omega$.   The result is shown in Fig.~\ref{fig:lingering}. We see that the particles which arrive at infinity with a high energy (i.e., the trajectories of high $\Omega$) hardly feel the presence of the Killing horizon. The particles  which arrive at infinity with a low energy (i.e., the trajectories for low $\Omega$) instead feel the presence of the Killing horizon acutely: they hover close to it for a considerable interval before escaping to infinity.

%------------------------------------------------------------------------------------------------------------------------------------------
%------------------------------------------------------------------------------------------------------------------------------------------

%Note that, as expected, these trajectories are \emph{not} geodesics due to the coupling with the {\ae}ther. The geodesics for this spacetime are similar to that of the Schwarzschild spacetime shown in figure \ref{schwarz}, peel off from the Killing horizon, and know nothing about the presence of the \aether\ or universal horizon as they are solely defined by the metric. 

%------------------------------------------------------------------------------------------------------------------------------------------
\section{Near-horizon physics}
%------------------------------------------------------------------------------------------------------------------------------------------
In this section we study the behavior of rays close to the Universal and Killing horizons. In Sec.~\ref{subsec:nearUH} the peel-off behavior of all rays, low or high $\Omega$, at the Universal horizon will let us define the surface gravity of the Universal horizon. 
In Sec.~\ref{subsec:nearKH} we attempt to define a surface gravity for the low-$\Omega$ rays close to the Killing horizon. 
In Sec.~\ref{subsec:lingeringKH} we attempt to quantify the extent to which the Killing horizon might reprocess the information in the low-$\Omega$ rays.

%------------------------------------------------------------------------------------------------------------------------------------------
\subsection{Near the Universal horizon}
%------------------------------------------------------------------------------------------------------------------------------------------
\label{subsec:nearUH}
%------------------------------------------------------------------------------------------------------------------------------------------
As seen in Fig.~\ref{fig:lingering}, there is a peeling behavior at the Universal horizon, therefore it is possible to associate with this behavior a notion of surface gravity. Surface gravity for Universal horizons has previously been considered in the literature   (see for example Refs.~\cite{Berglund:2012bu, Cropp:2013zxi}). 
Such derivations are non-trivial, given that the Universal horizon is qualitatively different from the Killing horizon, for which several alternative definitions of surface gravity all agree with each other (see Ref.~\cite{Cropp:2013zxi} for an exhaustive discussion). 
Let us stress that both Refs.~\cite{Berglund:2012bu} and~\cite{Cropp:2013zxi} use metric quantities and specific symmetries of spacetime to define the surface gravity of Universal horizons, and are in agreement with each other. 

However, in Ref.\cite{Berglund:2012fk}, the tunneling method~\cite{Parikh:1999mf} was used to claim that the Universal horizon should Hawking radiate at a particular temperature $T_{\UH}$. 
Given this temperature, a surface gravity can be associated to it in the usual way, $\kappa_{\rm thermal}={2\pi \, T}$. The surface gravity  thus defined, however, does not agree with the one derived in Refs.~\cite{Berglund:2012bu, Cropp:2013zxi}. 
As a consequence of this mismatch it was claimed in Ref.~\cite{Berglund:2012fk} that for Universal horizons the usual relation between surface gravity and temperature might break down.

Given this state of affairs, it is interesting to calculate the surface gravity in our framework, as the relevant quantity governing the observed peeling at the Universal horizon.
As we already understand the divergence at the Universal horizon, the calculation follows easily. We know that ${\rm v}_g \to \infty$ at the Universal horizon. 
This can be easily deduced from the fact that $\ks$ diverges there and that ${\rm v}_g=1+3\ell^2 \ks^2$ (see also Fig.~\ref{fig:k}).
Then from Eq.~\eqref{eq:dvdr} one can easily see that at the Universal horizon one gets
\begin{equation}
\frac{\d v}{\d r} = \frac{s^v}{s^r}.
\end{equation}
Thus the near-horizon peeling is the same as that calculated for the constant-$\tau$ slices at the Universal horizon, and it is independent of the specific form of dispersion relation  (as long as the dispersion relation is superluminal with unbounded group velocity, i.e., of the form as in Eq.~\eqref{eq:mod-disp} possibly with additional parity-odd terms). 
This can be confirmed by explicit computation (see appendix~\ref{gendispersion}). The surface gravity \eqref{eq:kappapeelUH} for the two exact solutions at hand is given by
\begin{equation}
\kappa_{c_{14}=0}=\frac{1}{2 r_{\UH}}\,\sqrt{\frac{2}{3(1-c_{13})}},
\qquad\hbox{and}\qquad
\kappa_{c_{123}=0}=\frac{1}{2 r_{\UH}} \,\sqrt{\frac{2-c_{14}}{2(1-c_{13})}},
\label{eq:kappaUH}
\end{equation}
respectively.

One can now easily check that the temperature calculated in Ref.~\cite{Berglund:2012fk}, and the above surface gravity, are indeed related in the standard way. 
While at the leading order we use the same dispersion relation as in Ref.~\cite{Berglund:2012fk}, we stress that our analysis shows that the peeling surface gravity of the Universal horizon is indeed universal, i.e., independent of the specific form of the superluminal dispersion relation. 
(An explicit demonstration is provided in Appendix~\ref{gendispersion}). 
This strongly suggests that it should be possible to carry out the tunneling calculation of Ref.~\cite{Berglund:2012fk} for general dispersion relations, and that the resulting temperature should be universal.

One might wonder why our analysis agrees with the temperature provided by the tunneling method, but does not agree with the surface gravity previously calculated in Ref.~\cite{Berglund:2012bu, Cropp:2013zxi}. 
Again the gist of the problem is that the methods of those references do not capture the role of \aether, and hence do not take into account the non-relativistic nature of the particle dynamics. On the contrary, the calculations presented in this paper rely on \aether\ in an essential way. 
Given that the Universal horizon is an \aether-dependent object, it is our  ``\aether\ sensitive" surface gravity which ends up being related to the temperature of the Universal horizon and has the physical meaning related to the peel-off behavior of ray trajectories.

In connection to this last comment one final remark is due. While it can be shown that the metric notion of surface gravity associated to geodesic peeling is equivalent to self-evidently covariant definitions~\cite{Cropp:2013zxi}, one might wonder if such alternative definitions are available for the peeling surface gravity associated to physical rays as discussed here. A natural candidate for a covariant definition of the surface gravity, which should match the peeling one, is the so called $\kappa_{\rm normal}$, which is equal (modulo a sign) to the normal derivative to the horizon of the redshift factor. For the case of the Universal horizon this simply takes the form $u^a\nabla_a (\chi^2)=-2\kappa^{\rm metric}_{\rm normal}$ and can be shown to be equal to the peeling surface gravity for geodesic rays~\cite{Cropp:2013zxi}. However, such a definition obviously does not capture the role of the {\ae}ther in the propagation of the physical rays. We will argue in Sec.~\ref{Hawkingrad} that a natural generalization of the redshift factor $\chi^2$ to our framework is given by $u\cdot\chi$ which is constant (actually zero) on the Universal horizon. Then, following the standard logical steps, one can recognize that the natural generalization of the above formula is $u^a\nabla_a (u\cdot\chi)=2\kappa_{\rm normal}$ evaluated on the Universal horizon. A straightforward calculation shows that using this definition yields the same value as in Eq.~\eqref{eq:kappaUH}. In fact, using the general form of the metric coefficients for spherically symmetric solutions given in Sec.~\ref{ae-bhs} one can show that $\kappa_{\rm normal}$ as defined by $u^a\nabla_a (u\cdot\chi)|_{UH}=2\kappa_{\rm normal}$ always equals the peeling-off surface gravity $\kappa_{UH}$ as defined by Eq.~\eqref{eq:kappapeelUH}. We therefore have a covariant expression for the surface gravity of the Universal horizon as defined by the peeling-off behavior of rays,
\begin{align}
\label{eq:covkappaUH}
\kappa_{UH}=\frac{1}{2} u^a\nabla_a (u\cdot\chi)\biggr\rvert_{UH}.
\end{align}
Using the Killing equation it can also be written as $\kappa_{UH}=\dfrac{1}{2}\chi \cdot a_u \biggr\rvert_{UH}$ where $a_u$ is the acceleration of the {\ae}ther, $a^b=u^c \nabla_c u^b$.

%\bigskip
%
%\hrule
%
%\medskip
%
%UPDATED UP TO THIS POINT
%
%\medskip
%
%\hrule
%
%\bigskip

%------------------------------------------------------------------------------------------------------------------------------------------
\subsection{Near the Killing horizon}
%------------------------------------------------------------------------------------------------------------------------------------------
\label{subsec:nearKH}
%------------------------------------------------------------------------------------------------------------------------------------------
In this section, we will work with the particular solution corresponding to $c_{123}=0$. As for the other solution, calculations become unpleasantly long and do not give any additional insight. 

At the Killing horizon, ($r=r_0+r_u$), the standard metric-determined surface gravity, (which can be found by any of the standard methods), is:
\begin{equation}
\kappa_{\KH, \mathrm{metric}}=\frac{r_0+2r_u}{4(r_0+r_u)^2}.
\end{equation}
Now at the Killing horizon, for the cubic dispersion relation
\begin{equation}
\ks|_{\KH}=\frac{[2(r_0+r_u)]^{1/3}\Omega^{1/3}}{(r_0+2r_u)^{1/3}\ell^{2/3}}, \qquad
%\end{equation}
%and
%\begin{equation}
\omega|_{\KH}=\frac{[2(r_0+r_u)]^{1/3}\Omega^{1/3}}{(r_0+2r_u)^{1/3}\ell^{2/3}}+\left[\frac{[2(r_0+r_u)]}{(r_0+2r_u)}\right]\Omega.
\label{eq:omegaOmega}
\end{equation}
While the group velocity is
\begin{equation}
{\rm v}_g|_{\KH}=1+\frac{3\ell^{2/3}\Omega^{2/3}(2(r_0+r_u)^{2/3})}{(r_0+2r_u)^2/3}=1+3\ell^{2/3}\Omega^{2/3}
\left[1+\left(\frac{r_0}{r_0+2r_u}\right)^{2/3}\right],
\end{equation}
which is still close to $1$ for small values of $\ell \Omega$, so for a dispersion relation only modified at high energies we are still almost relativistic until inside the Killing horizon. 

Can we define an approximate {\ae}ther-sensitive notion of surface gravity? We can certainly expand the ray trajectory in a Taylor series around the Killing horizon, trying to follow the standard peeling surface gravity calculation as given in Sec.~\ref{sec:standardpeeling}.
%\begin{equation}
%\left.\frac{\d r}{\d v}\right|_{\mathrm{out}}=\left.\frac{\d r}{\d v}\right|_{\KH}+\left.\frac{1}{2}\frac{\d}{\d r} \frac{\d r}{\d v}\right|_{\KH}(r-r_{\KH}) +\mathcal{O}(r-r_{\KH})^2.
%\end{equation}
However, as these rays have a non-relativistic dispersion relation the first term in this expansion \eqref{eq:standarddrdv} is no longer zero, and there is a net non-zero Killing-horizon-crossing velocity:
\begin{equation}
v_{\KH} = \left.\frac{\d r}{\d v}\right|_{\KH}.
\end{equation}
For the specific cubic dispersion relation considered above we have
% \begin{equation}
% \left.\frac{\d r}{\d v}\right|_{\KH}=\left(\frac{2(r_0+r_u)^2}{(r_0+2r_u)^2}+\frac{2^{4/3}(r_0+r_u)^{4/3}}{3(r_0+2r_u)^{4/3}(l\Omega)^{2/3}} \right).
% \end{equation}
\begin{equation}
\left.\frac{\d r}{\d v}\right|_{\KH}=\frac{3}{2}\,\frac{(r_0+2r_u)^2(2\ell\Omega)^{2/3}}{2(r_0+r_u)^{4/3}\left[2(r_0+2r_u)\right]^{2/3}+3(2\ell\Omega)^{2/3}(r_0+r_u)^{2/3}}.
\end{equation}
This is an outward-pointing, radial group velocity of the particle at the Killing horizon. Note that it has the correct relativistic limit, i.e., it vanishes as $\ell\to 0$. (The particular form above depends very much on the assumed cubic dispersion relation, but the fact that $v_{\KH} = ({\d r}/{\d v})|_{\KH}\neq 0$ is generic.)

In analogy with the usual treatment we can still use the second term of the Taylor expansion for defining a peeling surface gravity of the Killing horizon. The trick is to note that near the Killing horizon
\begin{align}
\left.\frac{\d r}{\d v}\right|_{\mathrm{out}}=\left.\frac{\d r}{\d v}\right|_{\KH}
+\left.\frac{\d}{\d r} \frac{\d r}{\d v}\right|_{\KH}(r-r_{\KH}) +\mathcal{O}(r-r_{\KH})^2.
\end{align}
Then if we compare two nearby trajectories $r_1(v)$ and $r_2(v)$, we see
\begin{align}
\left.\frac{\d (r_1-r_2)}{\d v}\right|_{\mathrm{out}}=\left.\frac{\d}{\d r} \frac{\d r}{\d v}\right|_{\KH}(r_1-r_2) 
+\mathcal{O}[(r_1-r_{\KH})^2, (r_2-r_{\KH})^2].
\end{align}
So this \emph{difference} certainly exhibits exponential peeling near the Killing horizon, with
\begin{equation}
\kappa_{\KH} = \frac{1}{2}\left.\frac{\d}{\d r} \frac{\d r}{\d v}\right|_{\KH}.
\end{equation}
For the specific cubic dispersion relation considered above we have
\begin{equation}
\kappa_{\KH}= \frac{r_0+2r_u}{4(r_0+r_u)^2}-{\frac{\left[2(r_0+2r_u)\right]^{1/3} \left(5r_0+3r_u\right){(\ell \Omega)^{2/3}}}{4\left(r_0+r_u \right)^{7/3}}} +\mathcal{O}(\Omega^{4/3}).
\label{eq:killingkappa}
\end{equation}
This is the closest notion we can construct to the usual metric surface gravity $\kappa_{\KH, \mathrm{metric}}$ at the Killing horizon once we are in the presence of modified dispersion relations.  
We see that it shows an $\Omega$ dependent correction to $\kappa_{\KH, \mathrm{metric}}$. For  $\ell \to 0$, this value agrees with the standard $\kappa_{\KH, \mathrm{metric}}$.  (Again, the particular form above depends very much on the assumed cubic dispersion relation, but the general features are generic.)

%Note
%\begin{equation}
%\kappa_{\KH, \mathrm{full}}=  \kappa_{\KH, \mathrm{metric}}
% -{\frac{[2(r_0+2r_u)]^{1/3} \left(5r_0+3r_u\right){\ell^{2/3}\Omega^{2/3}}}{\left(r_0+r_u \right)^{7/3}}} 
% +\mathcal{O}(\Omega^{2/3}).
%\end{equation}

We can also consider a quantitative comparison of the magnitude of the surface gravity at the Universal and Killing horizons. In particular, the ratio of Universal horizon surface gravity (see Eq.~\eqref{eq:kappaUH} for the $c_{123}=0$ solution), and Killing horizon surface gravity (Eq.~\eqref{eq:killingkappa}), is 
\begin{equation}
\frac{\kappa_{\UH}}{\kappa_{\KH}}=\frac{2(r_0+r_u)^2}{r_0^2} +\mathcal{O}(\Omega^{2/3}).  
\end{equation}
Hence, the surface gravity of the Universal horizon is higher than that of the Killing horizon. 

%------------------------------------------------------------------------------------------------------------------------------------------
\subsection{Lingering near the Killing horizon}
%------------------------------------------------------------------------------------------------------------------------------------------
\label{subsec:lingeringKH}
%------------------------------------------------------------------------------------------------------------------------------------------

% First let us get a estimate of the time as measured by the particle itself: $t_\mathrm{proper}=\frac{1}{\omega}$, and $\omega =k+l^2k^3$. We know the dependence of $k$ on $\omega$ at the Killing horizon, so
% \begin{equation}
% \omega=\frac{[2(r_0+r_u)]^{1/3}\Omega^{1/3}}{(r_0+2r_u)^{1/3}l^{2/3}}+\left[\frac{[2(r_0+r_u)]}{(r_0+2r_u)}\right]\Omega.
% \end{equation}

From the ray trajectories plotted in Fig.~\ref{fig:lingering} it is apparent that the low-$\Omega$ rays are strongly affected by the presence of the Killing horizon. 
These rays linger close to the Killing horizon before escaping out to infinity. It is conceivable that there is some sort of reprocessing going on in the vicinity of the Killing horizon. The ratio of the  time scale of lingering with respect to the intrinsic time scale associated to the ray seems like a good quantity to quantify the degree of reprocessing at the Killing horizon.  
In particular, one might try to compare how many ``ray cycles" at the Killing horizon, $\tau_{\mathrm{intrinsic}}=1/\omega |_{\KH}$, are contained in the lingering time, as this might give a good estimate of the extent to which the rays are significantly reprocessed. 
%The important concept here is how many cycles of frequency are undergone near the Killing horizon, and therefore the proper time is merely $\tau_{\mathrm{intrinsic}}=1/\omega |_{\KH}$

\begin{figure}[!htb]
\centering
\includegraphics[scale=1.0]{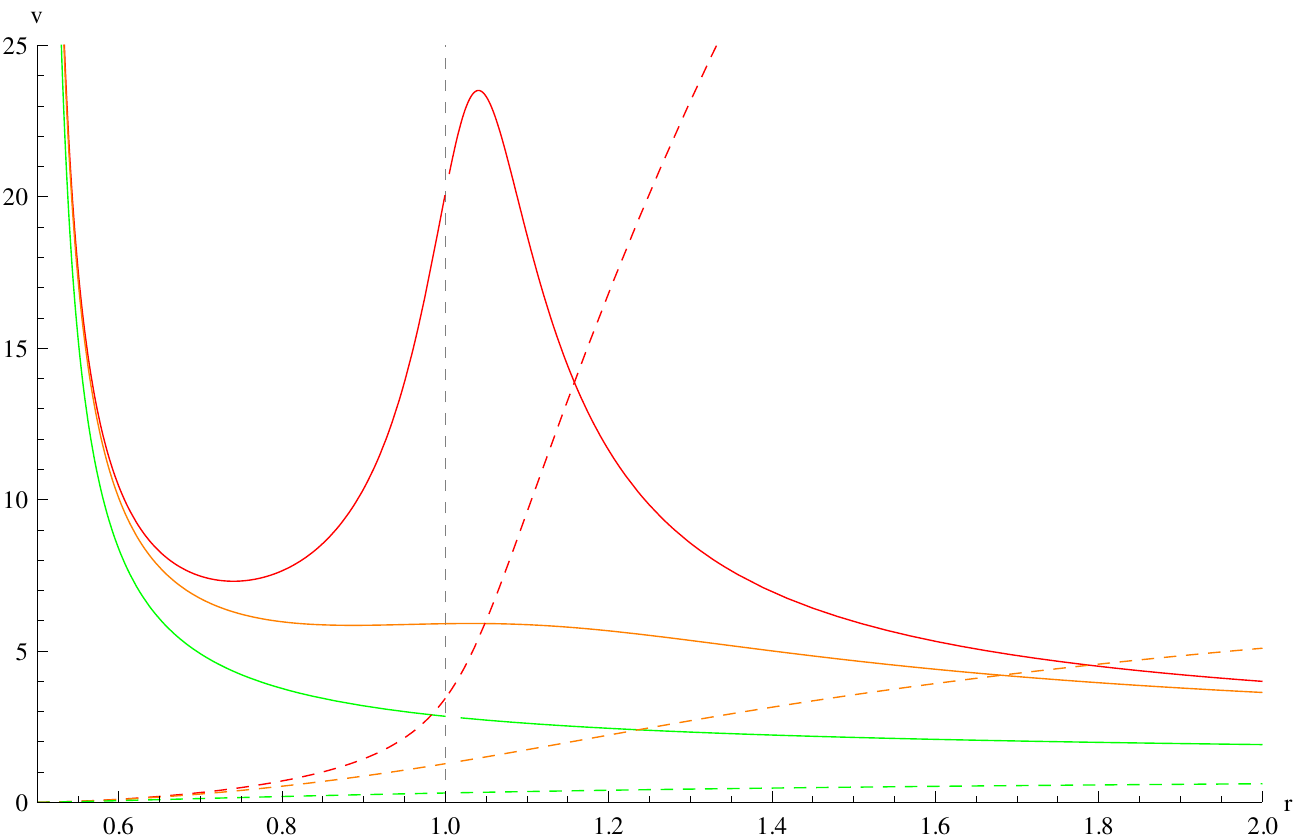}
\caption{$\ell/v$ (solid) and $\tau_{\mathrm{intrinsic}}$ (dashed), for $c_{123}=0$ solution, at $\Omega\ell=0.01$ (red), $\Omega\ell=0.1$ (orange), $\Omega\ell=1$ (green). Note the the peak is outside the Killing horizon}
\label{fig:times}
\end{figure}

An educated guess could then be to consider the ratio
\begin{equation}
{\mathcal{R}} = 
\frac{\tau_{\mathrm{linger}}}{\tau_{\mathrm{intrinsic}}}=  \,\ell \,\omega\,\left.\frac{\d v}{\d r}\right|_{\KH} ,
\end{equation}
where we have defined the lingering time as the ``width'' of the horizon, (for which $\ell$ is the simplest choice, though in the presence of modified dispersion relation a broadening of the horizon was found in, e.g., Ref.~\cite{Finazzi:2010yq}), times the crossing velocity. 
There is a problem with this, however. The lingering is \emph{outside} the Killing horizon, and the rays, even those with low energies, cross the horizon with a high coordinate velocity quite independently from how long they have lingered close to it. 
This can be seen in Fig.~\ref{fig:times}, which shows the time scales $1/\omega$ and $\ell/v$, (the time taken to go distance $\ell$ at instantaneous velocity $\d r/ \d v$). 
On the other hand, the previously defined $\kappa_{\KH}$ carries information about the concavity of the ray. If a particle has lingered it enters concave up, while those with high $\Omega$, will enter concave down.
%--------------------------------------------------------------------------------------------------------
%--------------------------------------------------------------------------------------------------------

%  Consider in particular figure \ref{fig:lingering}. If a particle has lingered it enters concave up, while those with high $\Omega$, will enter concave down. Therefore it seems that simple estimate of which particles linger is to see whether $\Omega$ is smaller than the zero concavity case (zero second derivative at just one point), given by
% \begin{equation}
% \Omega=\dots
% \end{equation}

Alternatively, we could consider the dispersion-dependent part of the Killing surface gravity in Eq.~\eqref{eq:killingkappa}, and write
\begin{equation}
\varkappa \equiv |\kappa_{\KH}-\kappa_{\KH, \mathrm{metric}}|={\frac{[2(r_0+2r_u)]^{1/3} \left(5r_0+3r_u\right){\ell^{2/3}\Omega^{2/3}}}{2\left(r_0+r_u \right)^{7/3}}}+ \mathcal{O}(\Omega^{4/3}),
\end{equation}
and then identify $\tau_\mathrm{linger}=1/\varkappa$. Note that this $\tau_\mathrm{linger}$ goes to zero for large $\Omega$, and becomes infinite as $\Omega$ goes to zero. 
Using Eq.~\eqref{eq:omegaOmega} for $\omega$, in the small $\Omega$ limit we now get
\begin{eqnarray}
{\mathcal{R}} = \frac{\tau_\mathrm{linger}}{\tau_\mathrm{intrinsic}}&=& \frac{\omega}{\varkappa}\approx
%\left(\frac{2\left(r_0+r_u \right)^{7/3}}{[2(r_0+2r_u)]^{1/3} \left(5r_0+3r_u\right){\ell^{2/3}\Omega^{2/3}}}\right)\left(\frac{[2(r_0+r_u)]^{1/3}\Omega^{1/3}}{(r_0+2r_u)^{1/3}\ell^{2/3}}\right) \nonumber \\ &=&
\frac{2(r_0+r_u)^{8/3}}{(r_0+2r_u)^{2/3}(5r_0+3r_u)} \frac{1}{\ell^{4/3}\Omega^{1/3}}.
\end{eqnarray}
So, rays with small $\Omega$  remain close enough to the Killing horizon for long enough to be significantly reprocessed. 

%{\bf Question:} {\sf Note
%\[
%\left.\frac{\d r}{\d v}\right|_{\KH}\neq 0.
%\]
%This is the ``velocity'' with which the dispersive ray trajectory crosses the Killing horizon. \\
%Shouldn't this be our a key ingredient in primary definition of lingering?\\
%Something like (horizon width)/(crossing velocity)?
%}
%
% \begin{eqnarray}
% \frac{\tau_{linger}}{\tau_{proper}} 
% &=&\left(2\,{\frac {{r_0}^{2}-{r_u}^{2}}{ \left(r_0+2r_u \right) ^{3}}}+
% {\frac {8}{9
% }}{\frac {(2({r_0+r_u}))^{1/3}r_0}{{{\Omega}}^{2/3} \left(r_0+2r_u\right)^{7/3}{l}^{2/3}}}
% +\frac{2}{9}{\frac{\left(r_0+r_u\right)^{2/3}{2}^{2/3}}{{{\Omega}}^{4/3}{l}^{4/3} \left(r_0+2r_u\right)^{5/3}
% }}\right)(r-r_{\KH})^2\nonumber \\
% &\times& \left(\frac{[2(r_0+r_u)]^{1/3}\Omega^{1/3}}{(r_0+2r_u)^{1/3}l^{2/3}}+\left[\frac{[2(r_0+r_u)]}{(r_0+2r_u)}\right]\Omega.\right)
% \end{eqnarray}

%------------------------------------------------------------------------------------------------------------------------------------------
%------------------------------------------------------------------------------------------------------------------------------------------

%------------------------------------------------------------------------------------------------------------------------------------------
\section{Hawking Radiation?}
%------------------------------------------------------------------------------------------------------------------------------------------
\label{Hawkingrad}
%------------------------------------------------------------------------------------------------------------------------------------------

What can we say about whether \emph{either} or \emph{both} of these horizons radiate? There are several basic ingredients necessary for Hawking radiation. However, studies related to the robustness of Hawking radiation show that these conditions only need to be valid at low energies. 
Firstly, we need a causal barrier enclosing an ergoregion, from which the energy needed to fuel Hawking radiation can be mined. 
In addition, we need an exponential peeling of rays from the causal barrier, which links the ingoing and the outgoing rays~\cite{Barcelo:2010xk}. Finally, we need an Unruh state, which is a vacuum state for the ingoing observer. \par
We have seen that the Universal horizon is a true causal barrier for particles of any energy. The peel-off behavior of the rays with modified dispersion relation is also universal, it does not depend on the exact dispersion relation as long as it is superluminal (see appendix \ref{gendispersion}). It is then tempting to associate a temperature with the Universal horizon determined by the surface-gravity in Eq.~\eqref{eq:kappaUH},
 \begin{equation}
T_{c_{14}=0}=\frac{1}{4\pi r_{\UH}}\,\sqrt{\frac{2}{3(1-c_{13})}},
\qquad\hbox{and}\qquad
T_{c_{123}=0}=\frac{1}{4\pi r_{\UH}} \,\sqrt{\frac{2-c_{14}}{2(1-c_{13})}}.
\end{equation}
This is in agreement with Ref.~\cite{Berglund:2012bu}, where the temperature associated with the Universal horizon was found using the Parikh-Wilczek tunneling formalism~\cite{Parikh:1999mf}. But it is not entirely clear as to where the energy for the Hawking radiation is mined from? 

In the case of Killing horizon, one has an intuitive picture of pair-creation close to the horizon: one particle has  positive energy while the second has negative energy. The positive energy particle escapes to infinity, while the negative energy particle which is classically not allowed to exist outside the horizon,  tunnels through the horizon and appears inside where it can now exist as a real particle because the Killing vector is spacelike there. 
From the point of view of the outside observer however, the negative energy has gone in, the black hole has lost mass which has been carried away as energy in the Hawking radiation. What is the analogous story in case of the Universal horizon? Is there an analog of the ergoregion behind the Universal horizon? \par
The relevant notion of energy in case of conventional Hawking radiation from a Killing horizon is the Killing energy. For a Universal horizon on the other hand, the relevant notion of energy is the energy measured in the {\ae}ther frame, $\omega=-k_a\, u^a$. 
Indeed, it is in the {\ae}ther frame that we have the modified dispersion relation, Eq.~\eqref{eq:mod-disp}. Now consider the pair-creation process right outside the Universal horizon. The particle created with positive $\omega$ escapes to infinity. As for the negative $\omega$ particle of the pair, from the equation for the conservation of energy  Eq.~\eqref{eq:conservation eqn} we have that for an ingoing mode with a negative energy outside the Universal horizon,
\begin{align}
\omega=\frac{-\Omega+\ks (\chi \cdot s)}{(\chi \cdot u)}.
\end{align}
We note that $\omega$ changes sign at the Universal horizon, with $\chi \cdot u$ now playing the role of the redshift factor $\chi^2$ in the standard calculation. This is so because $\chi \cdot u$ is negative outside, zero on, and positive inside the Universal horizon. Thus the negative energy particle of the pair after tunnelling through the Universal horizon appears inside with a postive $\omega$ and  propagates thereafter as a real particle. 
This suggests that the region behind the Universal horizon is analogous to ergoregion in the standard case and the picture that the pair-production close to horizon is responsible for the  Hawking radiation carries over to the Universal horizon. The only change being that the appropriate notion of energy is no longer the Killing energy but the energy measured in the {\ae}ther frame.  \par
There have been studies, mostly in the context of analogue spacetimes, which show that Hawking radiation from the Killing horizon is robust with respect to UV modifications of the dispersion relation~\cite{Unruh:2004zk,Barcelo:2008qe,Jacobson:1993hn}. 
In fact, the horizon does not even need to be an event horizon, but only some effective horizon~\cite{Barcelo:2006uw,Barcelo:2010xk}. This conclusion seems to be supported by Fig.~\ref{fig:lingering} where the low $\Omega$ rays seem to be peeling off the Killing horizon. 
However, there is a crucial difference. The cited studies do not have a Universal horizon whose presence will modify the boundary conditions for the modes. 
Once one realizes that there is in fact a Universal horizon one finds that, all the rays, irrespective of their energy, actually peel off the Universal horizon. 
The Killing horizon then seems to play the role of an efficient scattering surface for the low energy rays. The lesson is that the findings of analogue gravity can not be blindly applied to the case at hand. 
The reason, as discussed in Sec.~\ref{digression}, is simply that we do not (yet) have an analogue spacetime modeling the Universal horizon. Hence the intuition from the analogue gravity program must be exercised with caution. 

Finally, a crucial aspect of the derivation of Hawking radiation is the onset of an Unruh-type quantum vacuum state after the gravitational collapse. While there is evidence that Universal horizons form in spherically symmetric collapse~\cite{Saravani:2013kva}, we do not yet know the nature of the quantum vacuum state that is established at the end of this process. 
The experience based on black hole physics in general relativity~\cite{Barcelo:2007yk} would suggest that the quantum vacuum state for free falling observers (i.e., the Unruh state) should be established only at the Universal horizon because the universal and exact ``redshift"  associated with the peel-off behavior would  erase any leftover renormalized stress energy tensor exponentially fast at the horizon once it is formed. 
Furthermore, the evidence found in Ref.~\cite{Berglund:2012fk}, and in this paper, strongly suggests that thermal radiation will be produced at the Universal horizon and will escape the Killing horizon.  
This in turn suggests that the state experienced by the free falling observers at the Killing horizon would be non-generic, and would typically be different from the Unruh vacuum. 
To summarize, the presence of the Universal horizon inside the Killing horizon calls for a modification of the boundary conditions used for the derivation of the Hawking effect. A full calculation is needed to determine the spectrum observed at infinity. 
 
%------------------------------------------------------------------------------------------------------------------------------------------
\section{Discussion}
%------------------------------------------------------------------------------------------------------------------------------------------
\label{discussion}
%------------------------------------------------------------------------------------------------------------------------------------------

The main conclusion of the investigation undertaken in this paper is this: For the known Einstein-\AEther\ black holes, the relevant surface for Hawking radiation is \emph{not} the Killing horizon but the Universal horizon. 
This is the surface that admits an exact peeling behavior for ray trajectories. This is the surface for which the associated notion of surface gravity leads to the temperature obtained from tunneling methods. 
In this sense our work lends support to the findings of Refs.~\cite{Berglund:2012bu, Berglund:2012fk,Mohd:2013zca}, which were already pointing towards the Universal horizon being the surface relevant for thermal emission.

Furthermore, we have argued that the correct generalization of ergoregion in these  solutions is the region behind the Universal horizon, instead of the region behind the Killing horizon. 
We have also suggested that the appropriate vacuum state for  the free-falling observers must be realized at the Universal horizon instead of at the Killing horizon. 
A full calculation to address this issue would be very interesting. 

We have seen indications of the reprocessing of the low-energy particles at the Killing horizon. This seems to suggest that the thermal spectrum from the Universal horizon will be modified by the presence of the Killing horizon. 

To summarize, our work supports a consistent picture of thermal radiation from the Universal horizon, with an associated temperature determined by the peeling surface gravity \eqref{eq:kappaUH} in the standard way ($T=\kappa/2\pi$). 
The Killing horizon appears instead as a ``reprocessing/scattering surface" that  distorts the low-energy part of the original thermal spectrum in an energy and species (i.e., dispersion relation) dependent way. 

A shortcoming is that we cannot (with the current analysis) predict exactly what spectrum an observer at infinity will see from such a black hole. 
Further, there are several complications that, while not making the calculation impossible, will add an extra level of difficulty to determining such a spectrum. 
Even assuming that  Hawking radiation is produced at the Universal horizon one would still need to  include the role of the Killing horizon in reprocessing the outgoing low-energy modes. 

Nonetheless we feel that the clearer picture we have now, of the ergoregion behind the Universal horizon, the peeling-off behavior at the Universal horizon, and reprocessing of low-energy rays at the Killing horizon, shines new light on the thermodynamic character of black holes in Lorentz-violating theories. 
It is perhaps  too early to apply Sir Eddington's rule-of-thumb to these theories.

%------------------------------------------------------------------------------------------------------------------------------------------
\acknowledgments
%------------------------------------------------------------------------------------------------------------------------------------------

MV was supported by a Marsden Grant and by a James Cook Fellowship, both administered by the Royal Society of New Zealand. SL, AM and BC wish to thank Jishnu Bhattacharyya, Ted Jacobson, David Mattingly, and Thomas Sotiriou  for illuminating discussions and helpful comments on the manuscript.

%------------------------------------------------------------------------------------------------------------------------------------------
\appendix
%------------------------------------------------------------------------------------------------------------------------------------------
\section{General dispersion relations}
%------------------------------------------------------------------------------------------------------------------------------------------
\label{gendispersion}
%------------------------------------------------------------------------------------------------------------------------------------------

We have picked a superficially rather strange dispersion relation for our explicit numerical calculations. (It is certainly valid for low energies, simply being the expansion of $\omega^2=\ks^2+\ell^2 \ks^4$, where $\ks^4$ is generally expected to be the most relevant term at low energies). 
But we showed earlier that $\ks$ blows up near the Universal horizon, so how many of our conclusions can be carried over for more general dispersion relations?
Let us first look at the behavior near the Universal horizon. We still have the conservation equation
\begin{equation}
\label{appenomega}
\Omega =  \omega(\ks) |(\chi \cdot u)| - \ks |(\chi \cdot s)|.
\end{equation}
Again, on the Universal horizon we have $\chi \cdot u = 0$ and $\chi \cdot s = \|\chi \|$, we would naively but incorrectly get that
$ -\Omega = \ks \|\chi \|$, 
which is inconsistent because the RHS is positive while the LHS is negative. This argument holds just as well for a very large and relevant class of superluminal dispersion relations.
Let us now write 
\begin{equation}
\ks \sim \frac{a(r)}{|(\chi\cdot u)|^\gamma},
\end{equation}
near the Universal horizon, with $a(r)$ regular on the Universal horizon. Let us assume the dispersion relation has the form
\begin{equation}
\label{gendisp}
f(\ks)=\sum_{n=1}^{n=N} b_n \ks^{n}.
\end{equation}
Substituting this ansatz in the conservation equation, the finiteness of $\Omega$ implies that 
%
%The conservation equation becomes
%\begin{eqnarray}
%-\Omega &\sim&  b_Nk^{N} (\chi \cdot u) -  k(\chi \cdot s).\nonumber \\
%&\sim& \frac{b_n a(r)^{N}}{(\chi\cdot u)^{N\beta-1}}-\frac{\gamma(r) (\chi %\cdot s)}{(\chi\cdot u)^\gamma}
%\end{eqnarray}
%As $\Omega$ is always finite, these divergences must cancel, that is
\begin{equation}
N\gamma-1=\gamma; \qquad \hbox{so} \qquad \gamma=\frac{1}{N-1}; \qquad \hbox{and} \qquad a(r_\UH)^{N-1} = \frac{(\chi \cdot s)_\UH}{b_N}.
\end{equation}
How steeply $k$ diverges near the Universal horizon varies, but for superluminal dispersion relations of this form, it always will diverge. This would lead to a divergent group velocity at the Universal horizon. \par
In fact, we can generalize this argument even further to accomodate more general dispersion relations. It is now clear that what is really required for the peel-off behaviour and the universality of surface gravity is the divergence of the group velocity at the Universal horizon. Now, as the phase velocity is $v_{\mathrm{phase}}=\omega/k$, we can rearrange Eq.~\eqref{appenomega} as 
\begin{equation}
\Omega =  k_s [ v_{\mathrm{phase}} |\chi \cdot u| - |\chi \cdot s| ].
\end{equation}
If we assume the existence of outgoing modes at the universal horizon (meaning that this, rather than the Killing horizon, is the causal barrier), we necessarily have
\begin{equation}
v_{\mathrm{phase}} >  \frac{|\chi \cdot s|}{ |\chi \cdot u|},
\end{equation}
implying the phase velocity diverges at the Universal horizon. 
Now, using L'Hopital's rule, 
\begin{equation}
\lim_{k \to\infty} \frac{\omega(k)}{k} = \lim_{k \to \infty} \frac{[\d \omega(k)/\d k]}{[\d k/\d k]}=\lim_{k \to \infty} \frac{\d \omega(k)}{\d k} = \infty.
\end{equation}
So, (assuming sufficient smoothness in $\omega(k)$), the group velocity diverges at the Universal horizon. \par
Note other forms of superluminal dispersion exist, for instance extrapolating between two distinct limiting velocities. As Einstein-\Aether\ and \Horava-Lifshitz fundamentally violate Lorentz invariance, we do not expect to move between two Lorentzian regimes in this way, so dispersion relations of the form in Eq.~\eqref{gendisp} are the most relevant (also note for such rays the Universal horizon will not be the causal barrier).
Finally, using Eqs.~\eqref{eq:dvdr} and \eqref{eq:kappapeelUH}, we get a surface gravity which is universal and thus associates the universal temperature with the Universal horizon. 
\section{Conserved quantity}
\label{apdx:conservation}
In the geometric-optics approximation,  dynamics of the wavepacket is described as a point particle and is governed by some Hamiltonian,
\begin{align}
\label{eq:hamiltonian}
\mathcal{H}=\frac{1}{2}\mathcal{G}^{\alpha \beta}(x,k)\, {k}_{\alpha} {k}_{\beta}
\end{align}
where $k_\mu$ are the generalized momenta conjuate to the position coordinates $x^\mu$ of the particle (centre of the wave-packet), and $\mathcal{G}^{\alpha \beta}$ is a metric on the phase space. %$\mathcal{H}$ inherits reparametrization invariance from the diffeomorphism invariant of the underlying theory.
 Since the underlying field theory is diffeomorphism invariant, the $\mathcal{H}$ that descends from it is independent of the parameter along the trajectory of the particle. This implies that $\mathcal{H}$ is a constant on shell. Its value is the squared mass, which we take to be zero. Then Eq.~\eqref{eq:hamiltonian} is nothing but the dispersion relation \cite{Misner:1974qy}. \par
Now let the coefficients of the spacetime metric $g_{ab}$ in the Eddington-Finkelstein $\{v,r\}$ coordinates be independent of $v$, i.e., the coordinate vector field $\chi^a := \left(\frac{\partial}{\partial v}\right)^a$ is a Killing vector field. If the underlying dynamical field is Lie-dragged by $\chi$, the corresponding $\mathcal{H}$ has to respect this symmetry too, hence there is no explicit $v$-dependence in $\mathcal{H}$. Hamilton's equation $ \frac{\textstyle \rm{d}k_a}{\textstyle \rm{d}\lambda} = -\frac{\textstyle \partial \mathcal{H}}{\textstyle \partial x^a}$ then implies that $k_v$, the momentum conjugate to $v$, is a constant, i.e., $k_v := k_a \chi^a = - \Omega$, where $\Omega$ is a constant. \par
Suppose now that the dispersion relation is given in the {\ae}ther frame as $\omega = f(k_s)$, where $\omega=-k_a u^a$ and $k_s = k_a s^a$. The Hamiltonian is then $\mathcal{H}=\frac{1}{2}\left(\omega^2 - f(k_s)^2\right)$. Hamilton's equation for the evolution of the position is 
\begin{align}
\frac{\mathrm{d} x^a}{\mathrm{d} \lambda} &= \frac{\partial \mathrm{H}}{\partial k_a} \nonumber \\
 &= -\omega u^a - f(k_s) f'(k_s) s^a,
\end{align}
where $'$ denotes the derivative w.r.t the argument. Now noting that  $|f'(k_s)|$ is just the group velocity $\rm{v}_g$ we get, after dividing $\frac{\mathrm{d}r}{\mathrm{d}\lambda}$ and $\frac{d v}{\mathrm{d}\lambda}$, the equation describing the trajectory of the particle in the main text,
\begin{align}
\frac{\mathrm{d} r}{\mathrm{d} v} = \frac{ u^r \pm \mathrm{v}_g s^r}{ u^v \pm \mathrm{v}_g s^v},
\end{align}
 where $+(-)$ sign is for the outgoing(ingoing) particle.

%------------------------------------------------------------------------------------------------------------------------------------------

%------------------------------------------------------------------------------------------------------------------------------------------
\end{document}